\documentclass[12pt]{article}

\usepackage{xcolor}

\newcommand{\oi}{\hbox{[O$\,${\scriptsize I}]}}

\newcommand{\oiii}{\hbox{[O$\,${\scriptsize III}]}}

\newcommand{\nii}{\hbox{[N$\,${\scriptsize II}]}}
\newcommand{\sii}{\hbox{[S$\,${\scriptsize II}]}}
\newcommand{\ha}{\hbox{H$\alpha$}}
\newcommand{\hb}{\hbox{H$\beta$}}
\newcommand{\civ}{\hbox{C$\,${\scriptsize IV}}}

\newcommand{\kmsn}{{\rm km}\,{\rm s}^{-1}} 
\newcommand{\kms}{km s$^{-1}$}
\newcommand{\msun}{M$_{\odot}$} 
\newcommand{\eden}{cm$^{-3}$}

\newcommand{\ergs}{{\rm erg}\, {\rm s}^{-1} }
\newcommand{\myr}{M$_\odot$~yr$^{-1}$} 
\newcommand{\lbol}{{L_{\rm bol}}}

\newcommand{\htwo}{H$_{2}$}
\newcommand{\lsun}{\ensuremath{\mathrm{L}_{\odot}}}
\newcommand{\micron}{$\mu$m }
\newcommand\arcsec{\mbox{$^{\prime\prime}$}}%

\newcommand{\g}{\rm g}
\newcommand{\cm}{\rm cm}
\newcommand{\K}{\rm K}
\newcommand{\s}{\rm s}
\newcommand{\yr}{\rm yr}
\newcommand{\pc}{\rm pc}
\newcommand{\kpc}{\rm kpc}

\newcommand{\be}{\begin{equation}}
\newcommand{\ee}{\end{equation}}

\usepackage{scicite}
\usepackage{graphicx}
\usepackage{times}
\usepackage{amssymb}
\usepackage{amsmath}
\usepackage{url}
\usepackage[normalem]{ulem}

\newenvironment{sciabstract}{%
\begin{quote} \bf}
{\end{quote}}

\title{Probing the Physics of Dusty Outflows through Complex Organic Molecules in the Early Universe}

\author{Andrey Vayner,$^{1,17\ast}$ Tanio Díaz-Santos,$^{2}$  Carl D. Ferkinhoff,$^{3}$ Peter R. M. Eisenhardt,$^{4}$\and Daniel Stern,$^{4}$ Lee Armus,$^{1}$ Brandon S. Hensley,$^{4}$ Daniel Anglés-Alcázar,$^{5}$\and Roberto J. Assef,$^{6}$ Román Fernández Aranda,$^{2}$ Andrew W. Blain,$^{7}$  Hyunsung D. Jun,$^{8,9}$\and Norman W. Murray,$^{10}$ Shelley Wright,$^{11}$ Chao-Wei Tsai,$^{12,13}$ Thomas Lai,$^{1}$ \and Niranjan Chandra Roy,$^{5}$ Drew Brisbin,$^{6}$ Manuel Aravena,$^{6}$ Jorge González-López $^{14,15}$\and Guodong Li,$^{6,12}$ Mai Liao,$^{6}$ Devika Shobhana,$^{6}$ Jingwen Wu,$^{12}$ Dejene Zewdie$^{16}$
\and
\small{$^{1}$IPAC, California Institute of Technology, 1200 E. California Boulevard, Pasadena, 91125, CA, USA}\and
\small{$^{2}$Institute of Astrophysics, Foundation for Research and Technology–Hellas (FORTH), Heraklion, 70013, Greece}\and
\small{$^{3}$Winona State University, Winona, 55987, MN, USA}\and
\small{$^{4}$Jet Propulsion Laboratory, California Institute of Technology, 4800 Oak Grove Drive, Pasadena, 91109, CA, USA}\and
\small{$^{5}$Department of Physics, University of Connecticut, 196 Auditorium Road, U-3046, Storrs, 06269-304, CT, USA}\and
\small{$^{6}$Instituto de Estudios Astrof\'isicos, Facultad de Ingenier\'ia y Ciencias, Universidad Diego Portales,}\and
\small{ Av. Ej\'ercito Libertador 441, Santiago, Chile}\and
\small{$^{7}$School of Physics and Astronomy, University of Leicester, Leicester, LE1 7RH, UK}\and
\small{$^{8}$Department of Physics, Northwestern College,101 7th St SW, Orange City, 51041, IA USA}\and
\small{$^{9}$School of Physics, Korea Institute for Advanced Study,}\and
\small{85 Hoegiro, Dongdaemun-gu, Seoul, 02455, Republic of Korea}\and
\small{$^{10}$Canadian Institute for Theoretical Astrophysics, University of Toronto,}\and
\small{60 St. George Street, Toronto, ON M5S 3H8, Canada}\and
\small{$^{11}$Department of Astronomy and Astrophysics, UC San Diego, 9500 Gilman Drive La Jolla, CA 92093 USA}\and
\small{$^{12}$National Astronomical Observatories, Chinese Academy of Sciences, 20A Datun Road, Beijing, 100101, China}\and
\small{$^{13}$Institute for Frontiers in Astronomy and Astrophysics, Beijing Normal University, Beijing, 102206, China}\and
\small{$^{14}$Instituto de Astrof\'isica, Facultad de F\'isica,}\and
\small{Pontiﬁcia Universidad Cat\'olica de Chile, Santiago 7820436, Chile}\and
\small{$^{15}$Las Campanas Observatory, Carnegie Institution of Washington, Ra\'ul Bitr\'an 1200, La Serena, Chile}\and
\small{$^{16}$Centre for Space Research, North-West University, Potchefstroom, 2520, South Africa}\and
\small{$^{17}$Florida Gulf Coast University, 10501 FGCU Blvd. South, Fort Myers, 33965, FL, USA} \and
\small{$^\ast$To whom correspondence should be addressed; E-mail: avayner@fgcu.edu.}
}

\date{}

\begin{document} 


\baselineskip24pt 


\maketitle


\begin{sciabstract}

Galaxy-scale outflows are of critical importance for galaxy formation and evolution. Dust grains are the main sites for the formation of molecules needed for star formation but are also important for the acceleration of outflows that can remove the gas reservoir critical for stellar mass growth. Using the MIRI medium-resolution integral field spectrograph aboard the James Webb Space Telescope (JWST), we detect the 3.28 \micron\ aromatic and the 3.4 \micron\ aliphatic hydrocarbon dust features in absorption in a redshift 4.601 hot dust-obscured galaxy, blue-shifted by $\Delta$V=$-5250^{+276}_{-339}$ \kms\ from the systemic redshift of the galaxy. The extremely high velocity of the dust indicates that the wind was accelerated by radiation pressure from the central quasar. These results pave a novel way for probing the physics of dusty outflows in active galaxies at early cosmic time.





\end{sciabstract}




\section*{Introduction}
Outflows driven by active galactic nuclei (AGN) regulate the growth of the largest galaxies in the Universe \cite{choi15,Costa18,Mercedes-Feliz23,Wellons23}. They induce turbulence in the interstellar medium (ISM), which prolongs the cooling of gas, and expel gas and dust from galaxies, thus removing fuel necessary for future star formation. Quasar-driven outflows accelerate gas and dust to extreme velocities (above the escape velocity of the gravitational potential) and are thought to be the dominant way that the circumgalactic medium (CGM) of massive galaxies becomes enriched with dust and metals \cite{Travascio20,Choi20,Sanchez24}. \\

Outflows are inherently multi-phase with a range of temperatures and densities \cite{Fiore17,Rupke19,Richings18,Richings20,Ward24}. The colder phases often carry substantial mass, energy, and momentum \cite{Stone16,Lutz20,Vayner21mol}; hence, obtaining a full picture of the energetics in galaxy-scale outflows requires multi-wavelength observations. Molecules are commonly found in kiloparsec-scale outflows from local galaxies and from galaxies in the distant Universe, most commonly using either carbon-monoxide (CO) emission or hydroxyl (OH) absorption as a kinematics tracer \cite{Stone16,Vayner17,Spilker18,Lutz20,Vayner21mol,Salak24}. As molecules are unlikely to survive the powerful shocks that give rise to galaxy-scale outflows, those observed must be formed in-situ, a process that requires substantial amounts of dust \cite{Richings18}. Molecules form on the surfaces of dust particles; polycyclic aromatic hydrocarbons (PAHs) dominate the available dust surface area and are thus the natural sites of molecule formation \cite{Hensley23}. The presence of dust in the form of PAHs is crucial to explain the existence of smaller molecules in outflows. \\

In the nearby Universe, dusty outflows have been found using the \textit{Spitzer Infrared Space Telescope} through observations of PAH emission in the nearby starburst galaxy M82 \cite{Yamagishi12,Beirao15}, and in NGC 891 \cite{Rand08}. In the more distant Universe, PAH emission has been detected in both star-forming galaxies \cite{Yan05,Lutz05,Siana09,Riechers14} and those hosting AGN \cite{Chen24}. However, the 3.3 \micron\ aromatic and the 3.4 \micron\ aliphatic (chain-like) hydrocarbon features have only been individually detected in \textit{absorption} inside our own Milky Way galaxy and in a few nearby luminous infrared galaxies with very high nuclear obscuration (N$\rm_{H}\sim10^{24}~cm^{-2}$) \cite{Dartois04,Mason04,Chiar13,Hensley20}. The ensemble of the 3.4 \micron\ absorption feature has been detected in a few $z\sim2$ ULIRGs with \textit{Spitzer Infrared Space Telescope}; however, the low spectral resolution made it challenging to study the individual components of the complex \cite{Sajina09}. Aromatic or aliphatic hydrocarbon features have never been found directly in outflows outside the local ($z<0.1$) Universe, in either emission or absorption. However, JWST and the high spectral resolution of the Mid-Infrared Instrument (MIRI) allow for the study of the distribution, kinematics, and dynamics of the 3.3 \micron\ aromatic and 3.4 \micron\ aliphatic features to large cosmological distances \cite{Spilker23,Chen24}.




\section*{Target and observations}
Using the MIRI medium resolution spectroscopy mode \cite{Rieke15}, we targeted the most luminous infrared galaxy in the universe, WISE J224607.57-052635.0 (W2246-0526) at $z=4.601$\cite{diaz18} (look-back time of 12.5 Gyr), only 1.3 billion years after the Big Bang. The primary source of the bolometric luminosity in this hot dust-obscured galaxy (Hot DOG \cite{Eisenhardt12,Wu12}) is a heavily obscured quasar, with up to 300 magnitudes of visual extinction towards the nuclear region ($N_{H}\sim1.2\times10^{24}$~cm$^{-2}$;\cite{Fernandez-Aranda24}) due to dust and gas with an intrinsic bolometric luminosity of $3.5\times10^{14}$ \lsun \cite{Tsai18}. W2246-0526 is a known multiple merger system with several companion galaxies in a 40 kpc radius bridged by ionized gas and dust \cite{diaz18}, surrounded by a halo of diffuse emission from the circumgalactic medium. A powerful quasar-driven outflow is responsible for shocks on interstellar and circumgalactic medium scales \cite{Vayner25}. \\

\section*{Spectral modeling and results}
The spectrum of W2246-0526 at rest-frame near-infrared wavelengths (0.87-5.1 \micron) is dominated by thermal emission from hot ($>500$ K) dust. We detect absorption from the 3.3 \micron\ aromatic and from the 3.4 \micron\ aliphatic hydrocarbon features from both methyl (CH$_3$) and methylene (CH$_2$) groups attached to the aromatic skeletons \cite{Joblin96,Pendleton02,Yang16} consistent with dust grains that have a coating of aliphatic molecules (Figure \ref{fig:MRS_spectrum}). This is the first detection of the individual aliphatic C-H bond features outside the low redshift Universe ($z>0.5$) and the first detection of the 3.3\micron\ aromatic feature in absorption at high redshift ($z>2$), enabled by the sensitivity and spectral resolution afforded by JWST in the mid-infrared. We construct a model for the absorption profile by assuming each sub-feature is represented by a Gaussian. The 3.3\micron\ aromatic feature is represented by a single Gaussian, and the 3.4\micron\ aliphatic feature is fit with four Gaussian components. The optical depth and width of each aromatic and aliphatic hydrocarbon feature are free to vary in our spectral modeling, but the relative redshift of all features is fixed. We find that the absorption profiles are offset by - 5250 km\,s$^{-1}$ for both the aromatic and aliphatic features relative to the systemic redshift ($z=4.601$) of W2246-0526 (Figure \ref{fig:MRS_spectrum}). We find very similar widths for the features to what has been found in the Milky Way ISM towards the Galactic center Quintuplet Cluster and the nearby luminous blue hypergiant Cyg OB2-12 (Table \ref{tab:best_fit}, \cite{Chiar13,Hensley20}). The relative strength of each feature is also similar to the Milky Way. Our detection is consistent with dust entrained in a fast-flowing, uniform wind moving at -5250 \kms. Previous studies that detected molecular outflows found that the fastest velocities reach about 3000 \kms \cite{Stone16,Lutz20,Vayner21mol,Dan25}. Velocities $>3000$ \kms\ have not been seen before for a dusty or molecular outflow at any redshift, only previously found in hot gas traced through X-rays in type-1 quasars \cite{Reeves20} and in the warm ionized phase in broad-absorption line quasars, extremely red quasars, and hot dust-obscured galaxies \cite{perr19,Jun20,Choi20}.\\

Given the similar width and relative strength of the aromatic and aliphatic absorption spectral features in W2246-0526 and the sight line to Cyg OB2-12, the dust is likely similar to Galactic diffuse ISM. We, therefore, use the known ratio between the total hydrogen column density ($N_{H}$) towards Cyg OB2-12 and the measured aliphatic (3.4 \micron) optical depth to estimate the total hydrogen column density associated with the W2246-0526 absorbing material moving at -5250 \kms, assuming a dust to gas ratio similar to the Milky Way. Hensley et al. 2020 \cite{Hensley20} find $\rm N_H/\tau_{3.4}~=~4.5\rm \times10^{23}cm^{-2}$, this translates to a total hydrogen column density of $\rm \sim4.5\times10^{22}\,cm^{-2}$ (or mass column density $\Sigma\approx 0.15\,{\rm g}\,{\rm cm}^{-2}$) for the outflowing material traced through the hydrocarbon absorption. Using the relation between the hydrogen column density and V-band optical extinction for the Milky Way \cite{Guver09}, this corresponds to an $\rm A_V$ value of $\sim20$ for the outflowing dusty wind, which is lower than the total extinction towards the nuclear region ($\rm A_V\sim300$ \cite{Fernandez-Aranda24}). \\

\section*{Dynamics of the dusty outflow and discussion}
Given this large extinction, we expect that the optical and UV flux in our direction will be a small fraction of the bolometric luminosity. We estimate the optical luminosity to be $L_{opt}\approx 1.5\times10^{44}\, \ergs$, or about $10^{-4}$ of the infrared luminosity (see supplementary Figure \ref{fig:nuclear_NIRSPEC}). The UV luminosity is likely a few times higher, based on the results for Hot DOGs in Figure 3 of \cite{2015ApJ...805...90T}. The UV opacity of dusty gas is $\kappa_{UV}\approx 1000\,{\rm cm}^2\,{\rm g}^{-1}$ \cite{2003ARA&A..41..241D}, about 170 times the opacity in the infrared, $\kappa_d\approx 6\,{\rm cm}^2\,{\rm g}^{-1}$. Thus, the ratio of IR to UV radiative force is $F_{mIR}/F_{UV}\approx 17$ along our line of sight.

In fact, the radiative force ratio on the dusty wind is almost certainly larger, since the UV and optical radiation is very likely scattered, as evidenced by the large (10\%) polarization seen in other obscured quasars and Hot DOGs \cite{2018MNRAS.479.4936A, 2022ApJ...934..101A}. This scattered UV radiation will be absorbed before it can affect the bulk of the outflow, given the observed column density of the wind. We conclude that the infrared radiation accounts for the vast bulk of the acceleration due to radiation in the dusty outflow along our line of sight.

We estimate the location of the absorbing hydrocarbon material based on a few assumptions. The minimum radius at which the hydrocarbon dust could exist would be at the dust sublimation radius, where the dust interior is destroyed by the strong radiation field of the quasar:

\begin{equation}\label{eq:sublimation_r}
    \rm R_{sub} = \sqrt{\frac{L_{bol}}{4\pi\sigma_{SB}T_{sub}^{4}}} \approx 0.7~ \rm pc~ \bigg(\frac{L_{bol}}{10^{46}\,
    {\rm erg}\,{\rm s}^{-1}}
    \bigg)^{1/2}\bigg(\frac{T_{sub}}{1300\, K}\bigg)^{-2}.
\end{equation}


\noindent Assuming a quasar bolometric luminosity of $L_{bol}=1.3\times10^{48} \ergs$ and a dust sublimation temperature of $T_{sub}=1300$ K, consistent with observations of mid- to near-IR spectra of quasars \cite{2006ApJ...640..579G} leads to a minimum radius of $\approx 8$ pc. We see the 3.4 $\mu$m absorbing material against the mid-IR emission from dust grains, so the outflow is outside the dust sublimation radius. 

The maximum distance comes from assuming that the warm emitting large-grain dust is in thermal equilibrium with the quasar; the estimated effective radius for warm dust ($\sim$100 K) is $\sim$ 500 pc for W2246-0526 \cite{Fan16b,Fernandez-Aranda24}. Therefore, the absorbing material could potentially reside at a radius $\gtrsim$ 500 pc.


We can estimate the rate at which the material is outflowing by:
\begin{equation}
    \dot{M} = 4\pi\Omega N_{H} R \mu m_{p} v
\end{equation}

\noindent where $\Omega$ is the solid angle of the aromatic and aliphatic hydrocarbons, assumed to match the geometrical covering factor of the warm-dust ($f_c=0.92$)\cite{Fan16b}. $N_H$ is the hydrogen column density ($\rm \sim4.5\times10^{22}cm^{-2}$), $R$ is the radius (10-500 pc) where the outflow is currently located, $\mu\approx1.28$ is the atomic mass per proton assuming the outflow is neutral and atomic, $m_p$ is the proton mass, and $v$ is the velocity of the outflow, which we take to be 5250 \kms. Note that this assumes the absorbing material is a thin (nearly) spherical shell. Later we will model it as continuous but possibly clumpy wind, and show that the terminal velocity is likely closer to $10,000$ \kms. We obtain an outflow rate of 300-16,000 \myr, a kinetic power of

\begin{equation}
L_w\equiv\frac{\Omega}{2}\dot M v^2\approx\,(0.03-1.4)\times10^{47}{\rm erg}\,{\rm s}^{-1},
\end{equation}

which leads to a coupling efficiency to the quasar bolometric luminosity of 0.2-10\% (kinetic power divided by the quasar bolometric luminosity). However, given our theoretical modeling, we predict that the peak of the optical depth arises from gas around 20-30 pc, where the velocity is $\approx 5,000~\kmsn$, while the terminal velocity is higher, $v\approx 10,000$ \kms. This suggests that the most probable scenario is that the coupling efficiency is around 1\%. The upper end of the coupling efficiency is unusually high for quasar-driven outflows, about two times higher than the typical highest measured values \cite{Harrison18}. 

Given that the energetics are near the theoretical maximum for quasar-driven outflows, the absorbing material is unlikely to be located significantly beyond 500 pc, rather, it is likely to be located somewhere between the hot (500 K) and the warm (100 K) dust component. This is consistent with our theoretical prediction for the peak of the optical depth radius, where the absorption feature is predicted to be formed at 15-20 pc. Models in which outflows are driven by multiply scattered infrared radiation, with large covering fractions $f_c$ and large infrared optical depths $\tau_{IR}>10$ \cite{Bieri17} predict high coupling efficiencies ($\sim10\%$). In such models, these conditions hold until optically thin channels open as the outflow expands radially, allowing the infrared photons to stream freely.

In the JWST NIRSpec spectrum of W2246-0526, we see a factor of 2-5 decrease in the observed flux in all ionized lines beyond $\sim-5000$ \kms\ (Figure \ref{fig:nirspec-profiles}), coincident with the velocity of the fast outflowing dust. This likely indicates that the fastest (blueshifted by more than 5000 \kms) ionized gas is located in a region behind the dusty outflow along our line of sight and, therefore, attenuated. If we assume the high covering fraction obscuring material is nearly spherically symmetric surrounding the quasar accretion disk (Figure \ref{fig:schematic}), then the ionized emission likely comes from a high latitude angle, where there is less extinction from the dusty outflow than along sight-lines directly toward the black hole, but enough to cause the emission lines from the high velocity ionized gas to be partially absorbed.

We measure a total ionized gas outflowing mass of $(3\pm1)\times10^{5}$ \msun, an outflow rate of 300 $\pm$50 \myr\ and an outflow velocity of 9960 \kms (see Supplemental section). To differentiate between different driving mechanisms for the ionized outflow, we compare the momentum flux of the outflow ($\dot{P}_{outflow} = V_{outflow} \times  \dot{M}$) to the photon momentum flux of the quasar accretion disk ($\dot{P}_{Quasar} = L_{bol}/c$). Theoretical works predict a momentum flux boost ($\dot{P}_{outflow}/\dot{P}_{Quasar}$) that galactic scale outflows attain based on different driving mechanisms. In the case of outflows accelerated through radiation pressure on dust grains, theoretical works predict momentum boost values of 0.5-10 on pc to kpc scales \cite{Thompson15,Bieri17,Costa18}. We measure a momentum flux of $(0.5\pm0.2)\times\lbol/c$ for the ionized gas phase of the outflow (see Supplementary material) within 1.2 kpc from the quasar. This is consistent with the momentum boost expected for outflows driven by radiation pressure on dust grains, with dust playing a crucial role in accelerating the flow \cite{Thompson15,Bieri17,Ishibashi18,Costa18,Venanzi20,Barnes20}. High momentum flux ratios can also be achieved through energy-conserving outflows, where a hot, radiatively inefficient shock sweeps material out of the galaxy \cite{King10,fauc12b,Zubovas12,Zubovas14}. However, in these models, the interaction of the hot shock with the ambient interstellar medium could destroy the dust grains through further shocks, and the hydrocarbon dust would somehow have to reform in situ, which would require the survival of very dense ($10^{9}$ \eden) clouds in the wind.

If dust is responsible for accelerating the ionized outflow through radiation pressure, the sharp drop in the flux of the ionized gas emission profiles at the velocity of the $3.4\mu m$ absorption feature suggests that the hydrocarbon dust is associated with the very fast-moving, highly turbulent gas observed in the rest-frame optical emission lines of hydrogen and oxygen (Figure \ref{fig:nirspec-profiles}). If the difference in velocity between the ionized and dusty outflows is real, that would be consistent with model predictions that the dust in the outflow is expected to be moving slower than the hotter ionized gas phase, which can be further accelerated through high UV opacity \cite{Thompson15}. The outflow rates and energetics associated with the ionized and dusty outflow are shown in Table \ref{tab:nuclear_outflow_properties}.\\

The conditions for large infrared opacities are satisfied in the center of W2246-0526 based on recent ALMA observations \cite{Fernandez-Aranda24} that indicate a large central column density ($N_{H}>10^{24}$ cm$^{-2}$). Given the high dust opacity, the coupling of infrared photons to the dust is efficient, with likely multiple scattering of infrared photons off the dust grains before they escape the central region. The acceleration timescale is remarkably short under such conditions, and the dust and gas can be accelerated from velocities of a few hundred \kms\ to very large velocities ($>$ 5000 \kms) on timescales of thousands of years (\cite{Roth12}, also see supplemental material). The force from the radiation pressure can significantly exceed that of the gravitational potential of a $10^{9}$ \msun\ supermassive black hole (SMBH) and the gravitational potential of the galaxy, assuming a rotational velocity of 300 \kms \cite{Roth12}, hence launching an outflow out to significant (kpc-scale) distances (see the supplemental section for further calculations). The dust responsible for helping the acceleration likely originated in the circumnuclear region of W2246-0526 and formed in the ejecta of supernovae \cite{Bianchi07} or in slow, dense winds from late-type giant and supergiant stars \cite{Ferrarotti06}. The remarkable similarity between the spectral signature of the hydrocarbon absorption features in W2246-0526 and that of the Milky Way indicates that the stellar process responsible for forming aromatic and aliphatic hydrocarbon dust must be similar. Observations of additional dust features at longer wavelengths and other molecular features (e.g, \htwo, CO, H$_{2}$O) also present in the outflow are necessary to better understand the relative abundance of the hydrocarbon dust and whether it is similar to the Milky Way and other nearby systems. Interestingly, \textit{Herschel Space Observatory} observations of W2246-0526 suggest the presence of the 9.7\micron\ silicate absorption feature \cite{Tsai18} due to a flux deficit in the 70\micron\ \textit{Herschel} PACS filter. Sources with 3.3 \micron\ aromatic and 3.4\micron\ aliphatic features often exhibit strong silicate absorption \cite{Mason04}, hinting at a similar dust composition in W2246-0526 to nearby galaxies.

The rapid acceleration of the wind at smaller radii means the absorbing dust will be spread over a large range of velocities. The spread in velocity, in turn, translates to a spread in optical depth, diluting the absorption signal during the acceleration phase. At the same time, mass conservation dictates that the density drops as $1/R^2$, confining the absorption to velocities about a factor of two below the wind terminal velocity. This is likely the reason why we do not see a significant broadening in the lines compared to what we observe in the sight lines towards the Milky Way center. \\

The combination of large column densities and a strong source of infrared photons are the two most important conditions to drive the observed outflow. Because the acceleration happens through infrared photons, their energies are not sufficient to destroy the dust grains. Furthermore, the acceleration is very gentle with an acceleration rate of only 5$ \times10^{-3}~\rm cm~s^{-2}$, or about 200,000 times less than the acceleration due to gravity at the surface of the earth. Given that the photons involved in the acceleration are reprocessed infrared photons, the radius of acceleration has to be beyond the dust sublimation radius to allow for UV photons from the quasar accretion disk to be reprocessed into infrared photons by dust. Roth et al. \cite{Roth12} find that the majority of the dust acceleration by infrared photons happens just beyond the sublimation radius, and the outflow velocity has a very weak dependence on the radius after the dust has been accelerated to radii 2-3 times the size of the initial driving radius. Our best estimate is that for W2246-0526, the outflow was driven somewhere near 10 pc to explain the observed velocity, given the intense luminosity of our source. In our theoretical modeling, the outflow, when observed, is likely closer to 15-20 pc since that is where the largest ($\Delta\tau\sim0.1$) optical depth occurs for the hydrocarbon absorption. Based on the theoretical modeling, we estimate an outflow rate of $\approx 600$ \myr\ and coupling efficiency to quasar bolometric luminosity of $\approx 1$\%. Figure \ref{fig:schematic} is a schematic diagram of the inner kpc region, showing the multi-phase outflow driven by the central hyperluminous quasar.

\begin{figure}
    \centering
    \includegraphics[width=.9\linewidth]{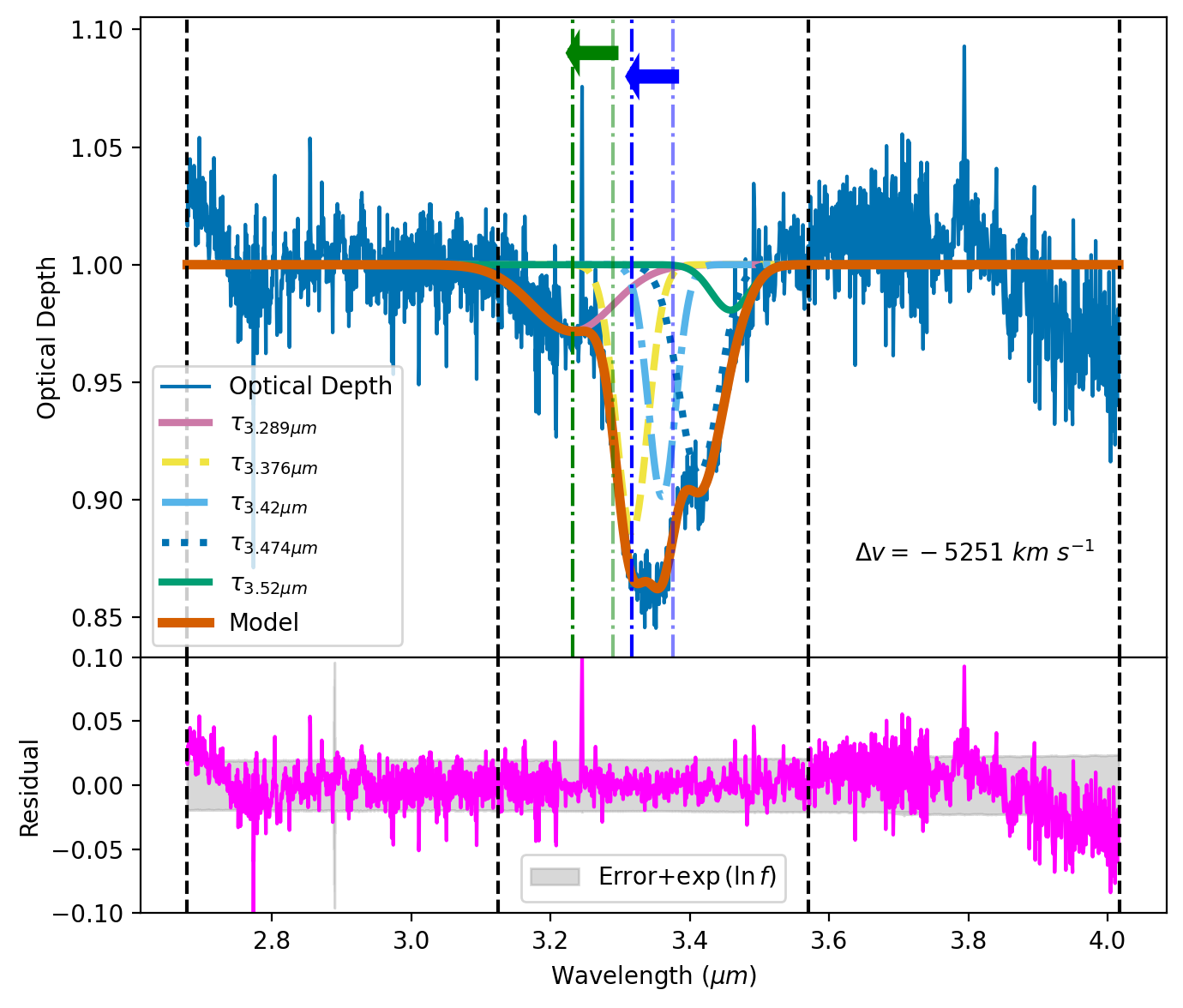}\\
    \caption{Rest-frame normalized MIRI MRS spectrum of the 3.3\micron\ aromatic and 3.4\micron\ aliphatic hydrocarbon complex in W2246-0526. The best-fit absorption model is shown in brown. Each Gaussian component represents a unique feature that makes up the 3.4 \micron\ aliphatic absorption complex with an additional Gaussian fit to the 3.3\micron\ aromatic absorption (Table \ref{tab:best_fit}). We measure a radial velocity offset of -5251 \kms\ in all the absorption features relative to systemic. The dashed-dot green and blue lines show the shift from systemic (light colors) to blueshifted (solid color) for the 3.289 \micron\ aromatic and 3.376 \micron\ aliphatic features, respectively, with the green/blue arrows showing the shift. The bottom panel shows the residuals of our best-fit model.}
    \label{fig:MRS_spectrum}
\end{figure}

\begin{figure}
    \centering
    \includegraphics[width=1.0\linewidth]{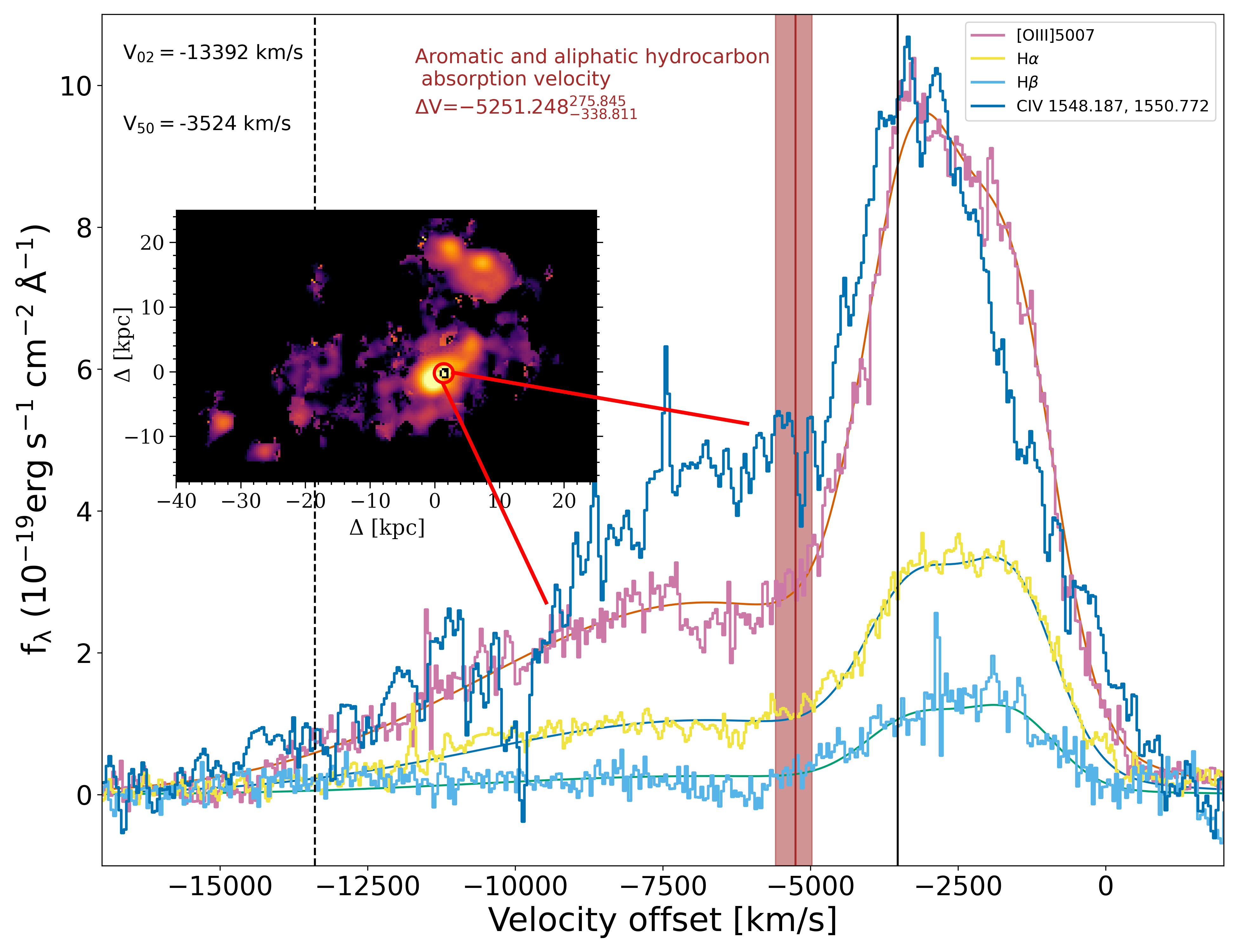}
    \caption{Line profiles of ionized emission from the outflow in the nuclear region of W2246-0526, marked with a red circle in the inserted figure showing the extended \ha\ emission in the W2246-0526 system, adapted from Vayner et al 2025 \cite{Vayner25}. Extremely broad and blueshifted emission is detected in \oiii, \ha\, and \hb, with maximum velocities of -13,400 \kms\ (V$_{02}$), where 0 \kms\ is systemic. The detection of \oiii\ indicates that this emission originates beyond the broad-line region of the quasar. We also overlay the \civ\ $\lambda \lambda$ 1548 \AA, 1550\AA\ emission line spectrum from rest-frame UV observations taken by the MUSE instrument, showing a similar profile to the optical lines. The solid black line shows the average velocity of the profile (V$_{50}$). The red line shows the velocity of the dusty outflow traced through aromatic and aliphatic hydrocarbon absorption.}
    \label{fig:nirspec-profiles}
\end{figure}

\begin{figure}
    \centering
    \includegraphics[width=0.8\linewidth]{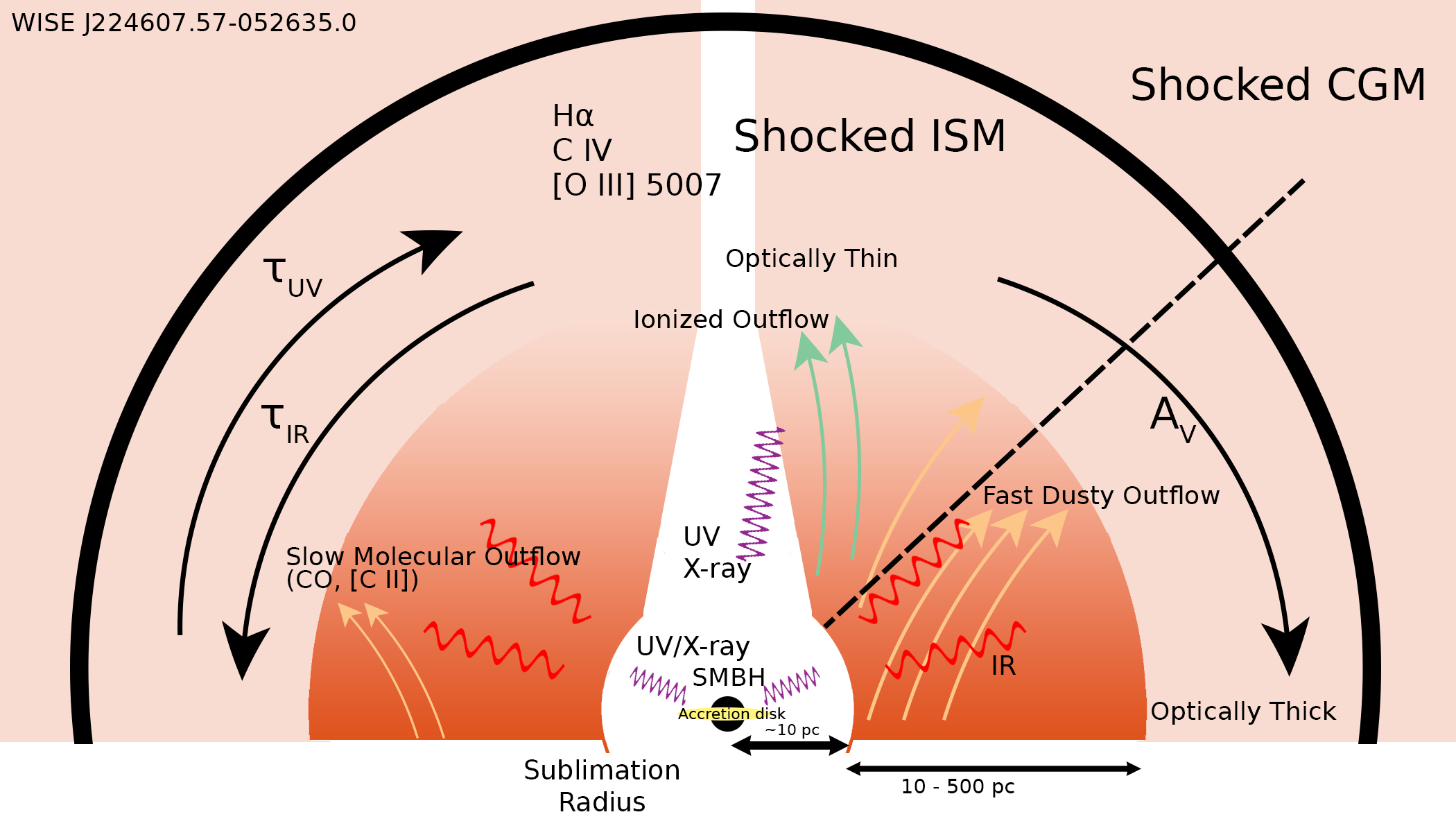}
    \caption{Schematic diagram of the inner kpc region of the W2246-0526 system. The supermassive black hole (SMBH) and the quasar accretion disk are located in the center. Surrounding the accretion disk is a dust-dominated region with a large covering solid angle around the quasar accretion disk and the broad-line region. High dust opacities, large column densities, and optically thick regions are located near the equatorial plane, with beige to orange colors representing an increase in these quantities. Past the dust sublimation radius, low to medium-latitude regions are dominated by infrared photons (red) formed through reprocessing UV photons originating from the quasar accretion disk. Light orange arrows mark the dusty outflow driven by radiation pressure from the infrared photons near the sublimation radius, and our line of sight crossing the path is marked with a dashed black line. Along the polar region, the column densities drop, allowing ionizing, UV, and optical photons to escape. The ionized outflow photo-ionized by the quasar likely escapes along this path. Part of the ionized emitting region is interior to the dusty outflow causing reddening of the emission, however not at a level to completely obscure it.}
    \label{fig:schematic}
\end{figure}

The detection of outflowing dust suggests that molecules can form and/or survive in the outflow, with aromatic and aliphatic hydrocarbons playing an important role as the smaller dust grains cover a large surface area. We predict that the fast molecular outflow should also be detectable in [CII]-158\,$\mu$m and other molecular gas tracers such as CO, OH, and H$_2$O emission and absorption; however, ALMA observations of these lines have not been made at a sufficiently blueshifted wavelength —  by more than 2500\,km\,s$^{-1}$ — to detect this emission or absorption. Observations of the dusty outflow in molecular gas tracers will help determine the amount of molecular gas in the outflow and will allow for a full multi-phase gas analysis to determine the distribution of mass and energetics among the different gas phases. The hydrocarbon absorption features may provide a direct measurement of the terminal velocity of the dust component of the multi-phase wind, thus a potential probe of the molecular conditions close to a powerful quasar, enabling new and clean measurements of accretion power and radiation intensity in the immediate environments of even extremely obscured powerful AGN. In emission, the 3.29 \micron feature has been detected out to $z=4.2$ with JWST \cite{Spilker23} in the lensed galaxy SPT0418-47. The associated aliphatic features at 3.4 \micron\ are undetected in SPT0418-47 as they are typically weaker in emission compared to the 3.29 \micron aromatic feature, consistent with what is generally seen in the local Universe. In the local Universe, in absorption, the aliphatic features are generally stronger compared to the aromatic features, similar to our detection in W2246-0526 (see Figure \ref{fig:MRS_spectrum}). The combined results from Spilker et al. \cite{Spilker23} and our results showcase that at least in these two systems in the distant universe, the relative strength of the aliphatic to aromatic features in emission and absorption is consistent with the overall situation in the local Universe. Recently, the detection of the 2175\AA\ UV extinction feature with JWST in star-forming galaxies out to $z\sim7$ \cite{Witstok23} indicates the presence of carbonaceous dust grains in place in the first billion years. The existence of large dust reservoirs at these early times indicates that quasar-driven outflows through radiation pressure on dust can happen well into the epoch of re-ionization. Other recent observations, of a quasar in the epoch of re-ionization at $z=7.5$ reveal an ionized outflow extending to 2 kpc\cite{Liu24}, the highest redshift object known, to date, with such an extended outflow. The energetics of the outflow may be consistent with radiation-pressure driving as the primary source behind the fast-moving ionized outflow, even at this early epoch.\\ 

We detect the 3.3 \micron\ aromatic and the 3.4 \micron\ aliphatic hydrocarbon features in absorption at $z=4.601$, the most distant direct detection of these features and the first detection of aromatic and aliphatic hydrocarbon features in absorption beyond the local Universe. The complex molecular features are blueshifted strongly, indicating the dusty outflow is likely driven by the central heavily obscured quasar through radiation pressure from reprocessed infrared photons. This supports the idea that efficient quasar-driven outflows can occur through radiation pressure on dust grains. Sufficient dust existing in powerful quasar-driven outflows indicates that molecules can form in the outflow, with aromatic and aliphatic hydrocarbon dust playing an important role. Given the large velocity observed for the dust in W2246-0526, quasars can be efficient sources of dust pollution in the early Universe into the interstellar, circumgalactic, and inter-galactic media \cite{Menard10,Chiang25}. Recently, hydrodynamical simulations show that indeed quasar-driven outflows such as the one in W2246-0526 can reach scales of several Mpc \cite{Borrow20,Gebhardt24}.

\begin{table}[]
    \centering
    \begin{tabular}{c c c c}
           & $\Delta$V=$-5250^{+276}_{-339}$ \kms &   & \\ 
         \hline
         Feature & Rest-Wavelength & $\tau$ & $\Delta \lambda$  \\
                 & \micron &   & \micron  \\
         \hline
         \hline
         CH($sp^2$) & 3.289 & $0.028^{+0.002}_{-0.002}$ & $0.058^{+0.005}_{-0.006}$\\
         CH$_3$($sp^3$) asym. & 3.376 & $0.114^{+0.005}_{-0.008}$ & $0.023^{+0.002}_{-0.002}$\\
         CH$_2$($sp^3$) asym.& 3.420 & $0.099^{+0.011}_{-0.012}$ & $0.020^{+0.002}_{-0.001}$\\
         CH$_2$($sp^3$) sym.& 3.474 & $0.089^{+0.004}_{-0.005}$ & $0.029_{-0.003}^{+0.005}$\\
         CH$_2$($sp^3$) sym.& 3.520 & $0.019_{-0.010}^{+0.010}$ & $0.028_{-0.005}^{+0.007}$\\

         \hline
    \end{tabular}
    \caption{Best fit parameters to the aromatic and aliphatic hydrocarbon features in the nuclear spectrum of W2246-0526. The rest-wavelength and the feature nomenclature follow Chiar et al. 2013 \cite{Chiar13}. The optical depth ($\tau$) is measured at the trough value while $\Delta\lambda$ is the dispersion in units of \micron\ for each individual aromoatic and aliphatic component.}
    \label{tab:best_fit}
\end{table}

\pagebreak

\bibliography{bib}

\bibliographystyle{Science}

\section*{Acknowledgments}
This work is based on observations made with the NASA/ESA/CSA James Webb Space Telescope. The data were obtained from the Mikulski Archive for Space Telescopes at the Space Telescope Science Institute, which is operated by the Association of Universities for Research in Astronomy, Inc., under NASA contract NAS 5-03127 for JWST. These observations are associated with program 1712.\\

\noindent Portions of this research were carried out at the Jet Propulsion Laboratory, California Institute of Technology, under a contract with the National Aeronautics and Space Administration.\\

\noindent RJA was supported by FONDECYT grant number 1231718 and by the ANID BASAL project FB210003.\\

\noindent DAA acknowledges support from NSF CAREER award AST-2442788, STScI JWST grants GO-01712.009-A, AR-04357.001-A, and AR-05366.005-A, an Alfred P. Sloan Research Fellowship, and Cottrell Scholar Award CS-CSA-2023-028 by the Research Corporation for Science Advancement. \\

\noindent Support for program \#JWST-GO-01712 was provided by NASA through a grant from the Space Telescope Science Institute, which is operated by the Association of Universities for Research in Astronomy, Inc., under NASA contract NAS 5-03127.\\

\noindent CF acknowledges support by NSF grant 1847892, and STScI grant JWST-GO01712.008-A\\

\noindent MA acknowledges support from ANID Basal Project FB210003 and ANID MILENIO NCN2024\_112\\

\noindent NWM acknowledges the support of the Natural Sciences and Engineering Research Council of Canada (NSERC), grant RGPIN-2023-04901. This work was performed in part at Aspen Center for Physics, which is supported by National Science Foundation grant PHY-2210452. \\

\noindent \textbf{Authors contributions:}
A. V. led the manuscript preparation, NIRSpec data reduction, and spectral analysis. T. D. -S led the telescope proposal and the MIRI data reduction. C. F. performed the MIRI spectral fitting analysis. B. H. and T. L. confirmed the Aromatic and Aliphatic features. D. S., L. A., D. A.-A., R. A., R. F.-A, A. B., H. J., S.W., C. W., N. C. R., D. B., M. A., J. G.-L., G. L., M. L., D. S., J. W., and D. Z., contributed to the discussion of the results. P. R. M. E. helped lead the discussions and provided significant assistance with editing the manuscript. N. W. M. provided the theoretical interpretation of the radiation pressure driving mechanism. \\

\noindent \textbf{Competing interests:} The authors declare that there are no competing interests. \\

\noindent \textbf{Data availability:}
The data is available at MAST, doi:10.17909/zydy-1534

\pagebreak
\section*{Supplementary materials}

\subsection*{Cosmology}
Throughout the paper, we use a standard $\Lambda$CDM cosmology \cite{Planck13} with $\Omega_{M}$ = 0.308, $\Omega_{\Lambda}$ = 0.692, and H$_{0}$ = 67.8 \kms\ Mpc$^{-1}$. At $z=4.601$, 1\arcsec\ corresponds to 6.690 kpc, the age of the universe was 1.301 Gyr, the luminosity distance is $4.33\times 10^{10}\,{\rm pc}$, and the look-back time is 12.496 Gyr. All emission and absorption lines are listed in vacuum wavelength.


\section{Data Reduction}
\subsection{MIRI MRS Observations and Data Reduction}\label{sec:obs_miri}

MIRI observations were taken on Jul 17, 2023 12:13:28 - Jul 17, 2023 15:03:41 (Proposal ID 1712) using the medium-resolution integral field unit mode. The instrument was set up to cover the full wavelength range (4.90-27.90 \micron) using the short, medium, and long grating setup with a 4-point dither pattern. The detector was set up to readout in the SLOWR1 mode with 31 non-destructive readouts for the short grating setup, 23 for the medium grating, and 15 for the long grating for a total exposure time per dither per grating setup of 740.58, 549.47 and 358.35 seconds. A background observation was taken at the end of the sequence with an offset to an empty sky region in a direction that placed the W2246-0526 in the center of the MIRI imager. The detector readout and exposure times for the background observations were matched to the science data. We did a two-point dither pattern during sky observations to help facilitate cosmic rays and artifact removal. Simultaneous MIRI imager frames were taken in the F560W, F770W, and F1000W filters.\\

Data reduction was done using the standard JWST pipeline version 1.16.1 and CRDS version 1322. The first stage of the pipeline performs regular infrared reduction steps, and the second stage of the pipeline flux calibrates the data, corrects for fringing effects, calibrates wavelengths, applies WCS, and performs sky subtraction. The third stage of the pipeline extracts the spectra from the two-dimensional frames into a datacube using a 3D ``drizzle" algorithm \cite{Law23}. We construct a ``master background" spectrum using the offset background observations and subtract it from the science data cube to remove the thermal background.

\subsection{NIRSpec IFU Observations and Data Reduction}\label{sec:obs_nirspec}
NIRSpec observations using the integral field unit were taken on July 16, 2023, 02:20:53 - 18:51:54 UT (Proposal ID 1712) using a combination of the G235H grating and F170LP filter and G395H grating and F290LP filter. The final resulting wavelength coverage is 1.66$-$3.15 \micron\ and 2.87–5.14 \micron, respectively, with a spectral resolution of 85-150 \kms \cite{Jakobsen22}.\\

NIRSpec IFU data reduction is outlined in detail in Vayner et al. 2025 \cite{Vayner25}. In short, we use the standard Space Telescope Science Institute (STScI) pipeline and run the ``Detector1Pipeline" and ``Spec2pipeline" to produce data cubes for each individual dither within a large 3$\times$2 mosaic. We then run custom routines designed to mitigate bad pixels, subtract background emission, and combine the data cubes into a single large data cube \cite{Vayner23b}.

\section{Analysis}
\subsection{MIRI Spectral fitting}\label{sec:fitting_miri}
\begin{figure}
    \centering
    \includegraphics[width=.8\linewidth]{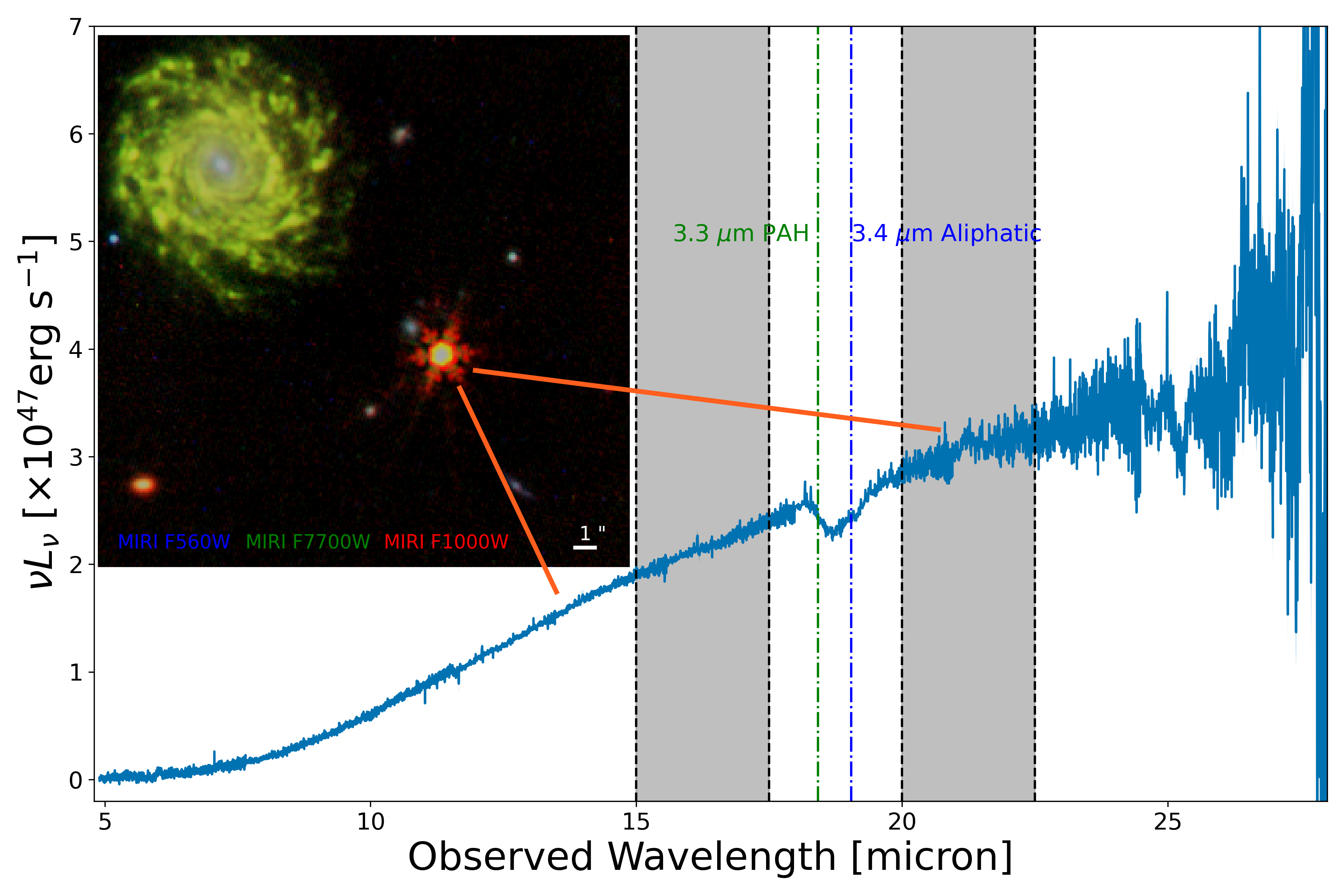}\\
    \caption{We present the total MIRI MRS spectrum in the observed wavelength from 4.92-28.6\micron, extracted over the point-source emission shown in the inserted 3-color MIRI imaging composite. The spectrum is dominated by hot dust emission from the central obscured quasar. The 3.3\micron\ aromatic and 3.4\micron\ aliphatic features are shown in dashed-dot green and blue lines at the systemic redshift of the source. The gray-shaded regions show the wavelength range used to estimate the continuum for fitting the aromatic and aliphatic absorptions. The bright galaxy in the top left of the insert is at $z=$0.092}
    \label{fig:total_MIRI_spec}
\end{figure}

\begin{figure}
    \centering
    \includegraphics[width=1.0\linewidth]{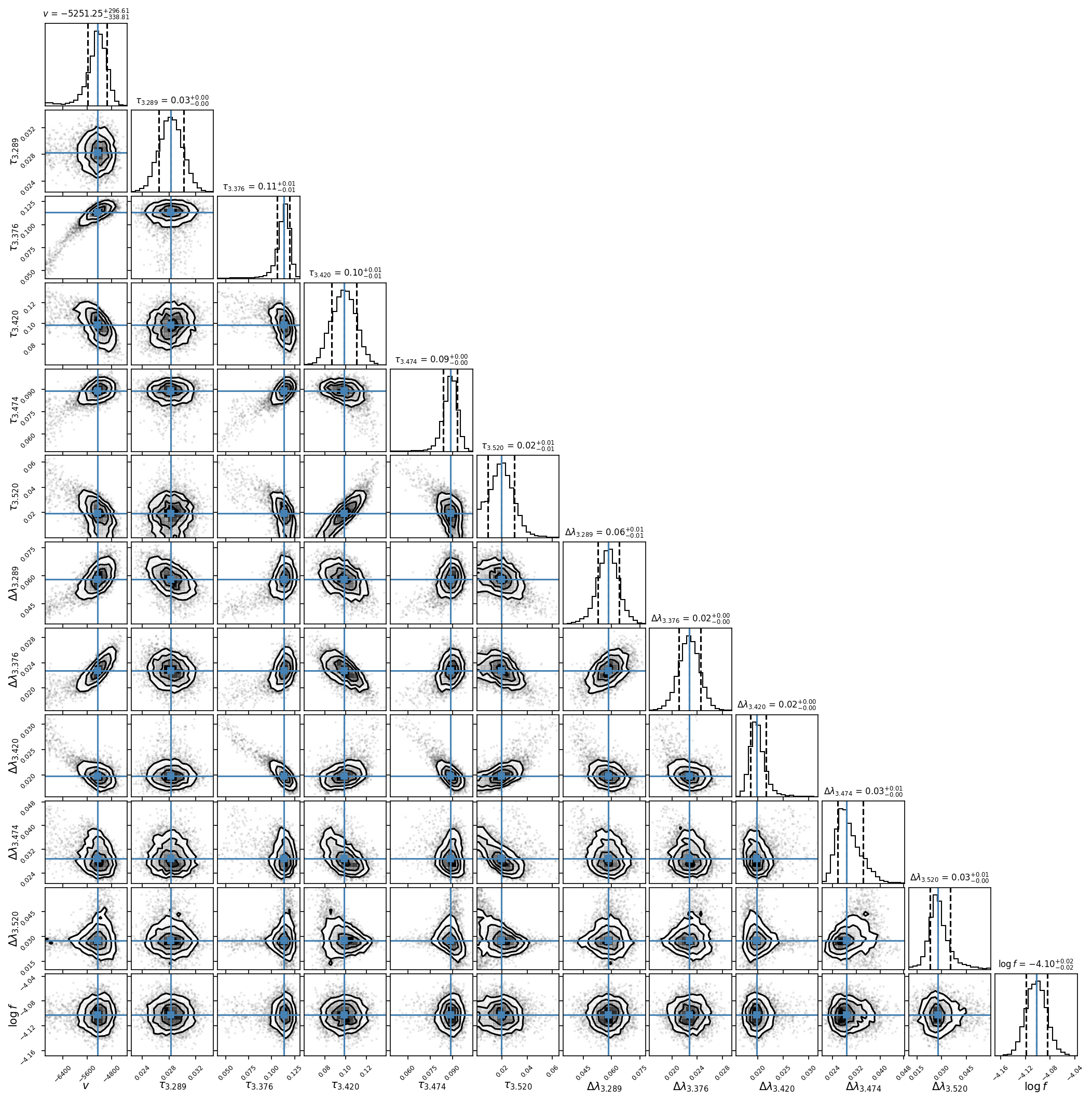}
    \caption{Corner plot from MCMC fitting to the normalized nuclear spectrum of W2246-0526.}
    \label{fig:corner}
\end{figure}

We extract a spectrum in a 0.25\arcsec\ radius circular aperture centered on the hot dust-obscured galaxy, selected to optimize the SNR across all wavelengths. The FWHM of the MIRI MRS PSF at the wavelength of aromatic and aliphatic features is 0.73\arcsec \cite{Law23}, the extracted aperture is about 68\% of the PSF size, as such, we apply an aperture correction to the spectrum when showcasing the total flux. We focus our analysis on the integrated spectrum (Figure \ref{fig:total_MIRI_spec}). We follow the steps of Chiar et al. \cite{Chiar13} in fitting the aromatic and aliphatic absorption features near 3.3 \micron\ and 3.4 \micron. We first normalize the spectrum by fitting a 5th-order polynomial to a line-free continuum window (shaded region in Figure \ref{fig:total_MIRI_spec}) around the absorption features. We divide the spectrum by the model to normalize the data. Using the rest-frame wavelength of the aromatic and aliphatic features from Chiar et al. \cite{Chiar13}, we fit each feature with a Gaussian absorption model; we limit the fitting range to 2.679-4.020 \micron\ in rest-frame wavelength to focus only on the absorption features. The width and optical depth of each feature are free to vary, but we fix the redshift of all the lines to be the same. We first run a maximum likelihood (ML) analysis on the absorption model with our initial guesses using the \texttt{scipy.optimize.minimize} function. We then run a full MCMC analysis \cite{Foreman-Mackey13} on the ML solution to explore the parameter space fully and estimate the uncertainties. We initialize walkers for each free parameter using the ML results and perturb each initial value using a random number generator. We then run MCMC using the \texttt{emcee}\cite{Foreman-Mackey13} package for 50,000 steps starting from the perturbed initial value. We use a thin value that is equal to the auto-correlation length, and we discard 10\% of the final chain. We also include an additional noise term ($\log(f)$) in our modeling to account for the under-reporting of errors that are known to exist in the MIRI-MRS pipeline\footnote{\url{https://jwst-docs.stsci.edu/known-issues-with-jwst-data/miri-known-issues/miri-mrs-known-issues#MIRIMRSKnownIssues-covarVariance,covariance,andDQarrays&gsc.tab=0}}. We list the best-fit parameters in Table \ref{tab:best_fit} and show the best-fit model in Figure \ref{fig:MRS_spectrum}. We present the corner plot from the MCMC analysis in Figure \ref{fig:corner}.

\subsection{NIRSpec Spectral fitting}\label{sec:fitting_nirspec}
Similar to the MIRI analysis, we only focus on the nuclear, spatially unresolved spectrum. We extract a spectrum using a 0.125\arcsec\ radius circle aperture centered on the hot dust-obscured galaxy. We select this aperture to minimize the inclusion of extended emission from the host galaxy while integrating over enough spaxels to average the spectral oscillating pattern due to the undersampling of the JWST PSF by the NIRSpec slicer \cite{Vayner23b}. We apply aperture correction to obtain the total point-source flux. Note that the aperture used is smaller than the MIRI because of the difference in the MIRI MRS and NIRSpec IFU PSFs. We fit \hb, \oi $\lambda$$\lambda$6300.304, 6363.776, \oiii $\lambda$$\lambda$4960.295, 5008.240, \nii\ $\lambda$$\lambda$6549.85, 6585.28, \ha, and \sii\ $\lambda$$\lambda$6718.29, 6732.67 simultaneously, the redshift and line dispersion are all held fixed to the \oiii\ line as it has the highest signal-to-noise ratio among the lines of interest. The lines model is convolved with the line-spread function of the NIRSpec IFS at each wavelength. The flux of \oiii $\lambda$$\lambda$4960.30, 5008.24 lines are held fixed to a 1:2.98 \cite{Storey99} line ratio, \nii $\lambda$$\lambda$ 6549.85, 6585.28 are held fixed to 1:2.96 \cite{Galavis97}, \oi
$\lambda$$\lambda$6300.304, 6363.776 are held fixed to a 3:1 ratio, and \sii\ $\lambda$$\lambda$6718.29, 6732.67 emission line ratios are allowed to vary between 0.5 and 1.5 \cite{Luridiana15}. The line ratio between \ha\ and \hb\ has to satisfy case B recombination conditions with a maximum ratio of 2.86 \cite{oste06}. We fit the nuclear spectrum with a number of broad ($V_{\sigma}>250$ \kms) and narrow ($V_{\sigma}<250$ \kms) Gaussian components. The continuum is fit with a 5th-order polynomial together with the emission lines. The line profiles are asymmetrical and are likely composed of many kinematic components among all the lines. Hence, we do not assign any meaning to the individual Gaussian components. The number of Gaussian components is selected to obtain similar line profiles among several emission lines in our best-fit model, especially \hb, \oiii\, and \ha. 

\begin{figure}
    \centering
    \includegraphics[width=1.0\linewidth]{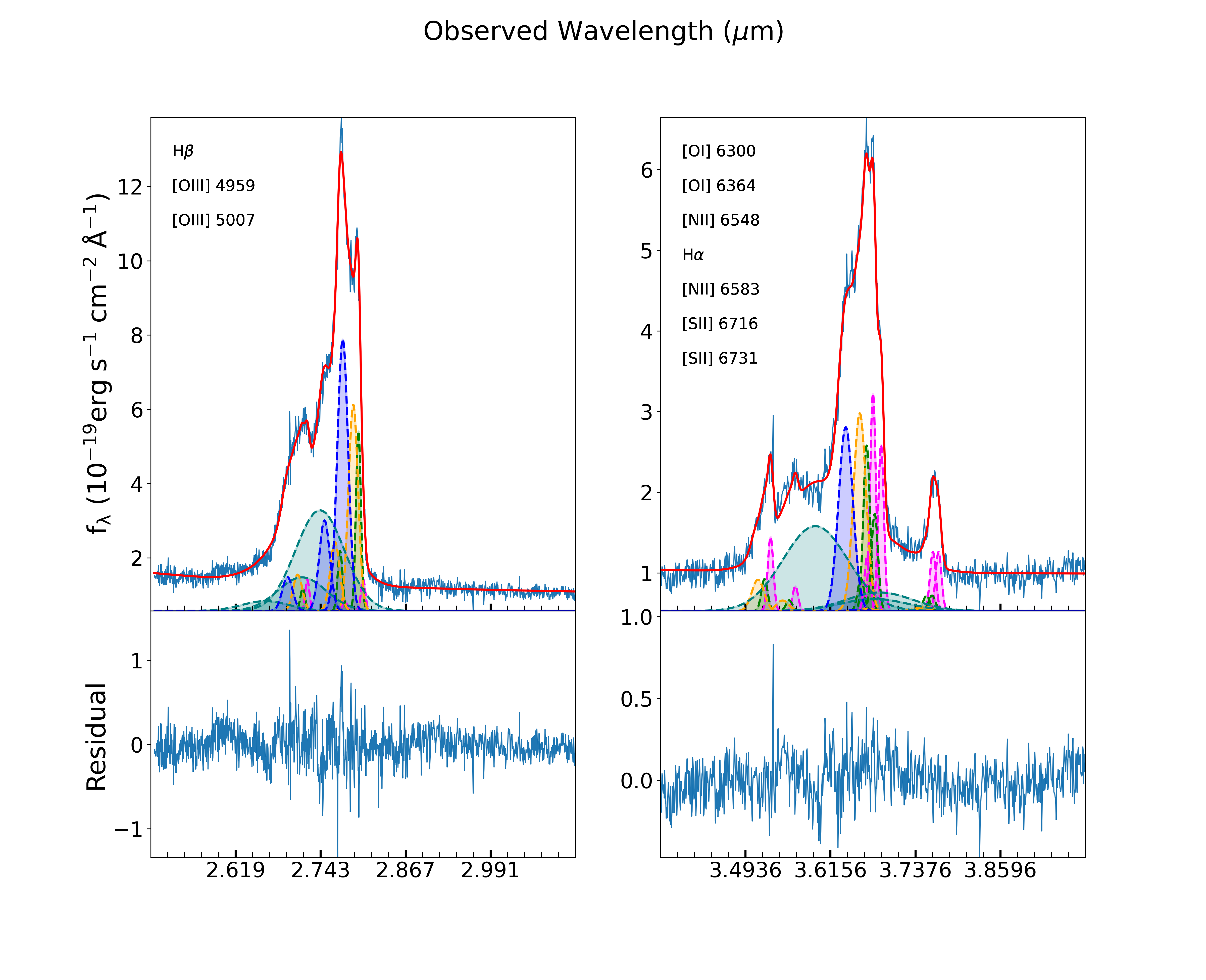}
    \caption{Top: best fit line and continuum model to the rest-frame optical nuclear spectrum of W2246-0526 focusing on the \oiii\ plus \hb\ (left) and \oi, \ha, \nii\ and \sii\ (right) line complex. The individual Gaussian for each line are shown in dashed lines matched in color for each kinematic component while the total continuum plus line model is shown in red. Bottom: residual emission line model and continuum spectrum. This figure showcases the individual kinematic components in different colors; in Figure \ref{fig:nirspec-profiles}, we show the profile of key ionized emission lines with a unique color. Figure reproduced from Vayner et al. 2025 \cite{Vayner25}}
    \label{fig:nuclear_NIRSPEC}
\end{figure}

The final best-fit line model consists of three broad ($V_{\sigma}>250$ \kms) and highly blueshifted ($>2000$ \kms) Gaussian models and two narrower components for the host galaxy near systemic velocity. We present the best fit model to the nuclear spectrum along with all the Gaussian components in Figure \ref{fig:nuclear_NIRSPEC} to both the \oiii\ plus \hb\ complex and to the \oi, \ha, \nii\ and \sii\ complex. In Figure \ref{fig:nirspec-profiles} we show the profiles of the individual best-fit lines, focusing on the \oiii, \ha\, and \hb\ lines. In addition, we overlay the \civ\ $\lambda \lambda$ 1548 \AA, 1550\AA\ emission line spectrum from rest-frame UV observations taken by the MUSE instrument (Shobhana et al. in-prep.), similarly to the NIRSpec data we extract the rest-frame UV spectrum covering the \civ\ line over an aperture solely encompassing the nuclear emission. The \civ\ profile shows a similar shape to the rest-frame optical emission lines; however, it requires no fitting and subtraction of any contaminating lines as the \civ\ emission comes from a region relatively clear of other strongly emitting lines.

To gauge the maximum and average velocities associated with the broad and blueshifted line emission profile, we use non-parametric analysis using the normalized velocity distribution function:

\begin{equation} 
    F(v)=\int\limits_{-\infty}^{v}f(v')dv'/\int\limits_{-\infty}^{\infty}f(v')dv'
    \label{eq:NCVD}
\end{equation}

\noindent We measure the maximum blue-shifted velocity ($V_{02}$) by setting Equation \ref{eq:NCVD} to 0.02 and solving for the velocity value equal to where 2\% of the flux is integrated on the blueshifted side of the broad velocity profile. In addition we measure the mean velocity ($V_{50}$) and the width where we integrate 90\% of the broad velocity profile ($W_{90}=V_{95}-V_{05}$). These values are listed in Figure \ref{fig:nirspec-profiles}. We see outflowing gas with incredibly high velocities, with maximum values reaching 13,000 \kms\ and an average velocity of -3500 \kms. No realistic galaxy gravitational potential can contain such fast and turbulent gas.\\

\subsection{Mass loss rate in the ionized wind}\label{sec Halpha}
In this subsection we give a rough calculation of the mass loss rate associated with the ionized gas that emits the broad H$_\alpha$ line. We refer to Figure \ref{fig:schematic}, in which we assume that the ionized outflow emerges from the throat of the torus opening (both above and below the torus, although the latter is not shown in the figure). We noted in the main text that the covering factor of the optically thick material was $f_c\approx0.92$, so the maximum potential covering factor of the UV radiation not blocked by the torus is $f_{UV}\equiv 1-f_c\approx 0.08$. We further assume that the [O III] emission is produced in an outflow emerging from or near the opening of the torus, rather than from emission interior to the torus and then being scattered. This allows us to get an upper limit on the mass loss rate.  We now show that $f_{UV}$ is much smaller than this upper limit.

We start by noting that the number of ionizing photons emitted by the quasar is 
\be
Q\approx \frac{1}{4}\frac{L_{\rm bol}}{\langle h\nu\rangle}\approx 1\times 10^{58}~\s^{-1},
\ee
where $\langle h\nu\rangle\approx 30\,{\rm eV}$ is the average energy of the ionizing photons, and we have assumed that about a quarter of the bolometric luminosity comes out beyond the Lyman edge.
The number density of ionizing photons is
\be
n_\gamma=\frac{Q}{4\pi r^2 c}\approx 3\times10^7\left(\frac{r}{10\,\pc}\right)^{-2}\,\cm^{-3}.
\ee

If every ionizing photon were absorbed by a hydrogen atom, the resulting H$_\alpha$ line luminosity would be 
\be
L_{{\rm H}_\alpha}\approx \frac{1}{2}h\nu_\alpha Q\approx1\times10^{46}\,\ergs,
\ee
where the factor of $1/2$ comes from the average number of H$\alpha$ photons emitted for every ionizing photon absorbed by a hydrogen atom. This is much larger than the observed value, $L^{\rm obs}_{{\rm H}_\alpha}\approx4.5\times10^{43}\,\ergs$, so we infer that the covering factor of the gas producing the H$_\alpha$ emission is 
\be
f_{{\rm H}_\alpha}\approx 4\times10^{-3}.
\ee

This inferred covering factor is much smaller than the covering factor of the opening in the torus, by a factor  $f_{{\rm H}_\alpha}/f_{UV}\approx 0.06$, suggesting that the ionized outflow emitting the $H_\alpha$ and by extension the $[O III]$ line, trace a hollow cone. This type of geometry is commonly seen in nearby AGN, e.g., \cite{2000ApJ...532L.101C, 2019ApJ...870...37D}. However, most locally observed hollow cones have a larger fraction of the cone that contains gas that emits the narrow lines; a similar fraction on W2246-0526 would over-produce the H$_\alpha$ emission. In local objects the narrow line gas is clearly dusty; in dusty gas exposed to a large flux of ionizing radiation, the neutral hydrogen fraction is small enough that dust grains can absorb a substantial fraction of the ionizing radiation.

This suggests that the source of material for the outflow is dusty gas in or near the throat of the torus. This gas could be driven out by ultraviolet radiation hitting grains in the dusty gas, or by a hot shocked wind emerging from well inside the torus, or a combination of the two. The presence of a hot shocked wind, which can drive lower temperature, UV or optically line emitting gas to high velocity, is commonly inferred to exist in quasars, e.g., \cite{2012MNRAS.425..605F}, and we will assume that such a wind exists in W2246-0526.

We further assume that this hot gas pressurizes the cooler ($\sim 10^4\,{\rm K}$) gas emitting the optical lines. The hot gas is produced by a broad absorption line (BAL) or continuum wind with a density given by
\be\label{equation:hot gas density}
n_h=\frac{\eta_{\rm bal}f_{\rm bal}\lbol}{\pi r^2cm_p v_{\rm bal}^2}
\approx 1
\left(\frac{\eta_{\rm bal}}{0.01}\right)
\left(\frac{r}{10\,\pc}\right)^{-2}
\left(\frac{v_{\rm bal}}{30,000\,\kmsn}\right)
\,\cm^{-3}.
\ee
In this expression $n_h$ is the number density of the post-shock hot gas, $\eta_{\rm bal}\approx 0.01$ is the fraction of the UV radiation scattered by lines (pre-shock), $f_{\rm bal}\approx0.2$ is the covering factor of the BAL wind, and $v_{\rm bal}\approx 30,000\,\kmsn$ is a characteristic velocity of the wind. For a continuum driven wind $\eta_{\rm bal}\approx 1$, in which case the subscript is a misnomer.  The post-shock temperature of the wind is 
\be
T_h=\frac{3\mu m_p}{16k_b}v_{\rm bal}^2\approx10^{10}\left(\frac{v_{\rm bal}}{30,000\,\kmsn}\right)^2\, \K.
\ee

Assuming pressure equilibrium between the hot gas and the line emitting gas, the latter has a density
\be
n_{\rm cl}=\frac{3\eta_{\rm bal}f_{\rm bal}\lbol}{32\pi r^2 c\, k_b T_{cl}}
\approx 10^6
\left(\frac{\eta_{\rm bal}}{0.01}\right)
\left(\frac{r}{10\,\pc}\right)^{-2}
\left(\frac{\lbol}{10^{48}\ergs}\right)\left(\frac{T_{\rm cl}}{2\times10^4\,\K}\right)^{-1}
\,\cm^{-3},
\ee
where $k_b\approx 1.38\times10^{-16}\,\ergs\,\K^{-1}$ is Boltzman's constant, and $T_{\rm cl}$ is the temperature of the H$_\alpha$ emitting gas clump.
This is slightly larger than the critical density of the \oiii\ 5007 line of $n_{\rm crit\ [OIII]}\approx7\times10^5\,\cm^{-3}$ \cite{Baskin05}.

We can use this result to estimate the ionization parameter
\be
U\equiv \frac{n_\gamma}{n_{cl}}\approx 30\left(\frac{\eta_{\rm bal}}{0.01}\right)^{-1}.
\ee

We estimate the ionized gas mass following the methodology of \cite{oste06}, assuming case B recombination:

\begin{equation}\label{equation:ionized_gas_mass}
    M_{Ionized} = \frac{\mu m_{p}L_{H\alpha}}{4\pi j_{H_{\alpha}}\bar n_{\rm cl}}
    \approx6\times10^{38}\left(\frac{\bar n_e}{9\times10^5\,\cm^{-3}}\right)^{-1}\,\g,
\end{equation}
where $\mu=1.4$ is the mean molecular weight of ionized gas, and $\rm L_{H\alpha}$ and $\rm \bar n_{\rm cl}$ are \ha\ luminosity and average electron density, respectively. The quantity $4\pi j_{H_{\alpha}}=h\nu_{H_\alpha}\alpha^{eff}_{H_\alpha}$ is the \ha\ emissitivity, constrained to be within (1.8-3.53)$\times10^{-25}\rm~erg~cm^{-3}~s^{-1}$ for a gas temperature of (1-2)$\times10^{4}$ K \cite{oste06, Luridiana15}.
Using $4\pi j_{H_\alpha}=2\times10^{-25}\,\ergs\,\cm^{-3}$, we estimate the outflow rate to be
\begin{equation}\label{equation:outflow-simple}
    \dot{M}{_{Ionized}}=\frac{M_{ionized}v_{out}}{R}\approx 300\left(\frac{R}{10\pc}\right)^{-1}\left(\frac{\bar n_e}{9\times10^5\,\cm^{-3}}\right)^{-1}\,M_\odot\,\yr^{-1},
\end{equation}
where $M_{ionized}$ is the ionized gas mass, $v_{out}$ is the outflow velocity taken to be as the $V_{05}\approx 10,000\,\kmsn$ value from the broad blueshifted kinematic components and $R$ is the radius of the outflow.

The associated momentum flux and kinetic luminosity of the outflow are listed in Table \ref{tab:nuclear_outflow_properties}.


We note that if $A_v\approx23$ (as measured in the dusty outflow) was the true optical depth over the entire source, no UV or optical radiation would emerge. However, we see some UV continuum and line emission, the latter including \civ\ 1550 emission (Figure \ref{fig:nirspec-profiles}), with a substantial velocity width. We infer that there is a moderately high ionization outflow and that the gas in that outflow is visible at Earth.

The \civ\ line shows some absorption around $5000$ \kms, consistent with absorption by the 3.4 \micron\ absorption feature seen in the continuum (Figure \ref{fig:nirspec-profiles}). However, the UV continuum is not absorbed by the \civ\ line, indicating that the line emitting gas is not flowing toward us and that we do not have a direct line of sight toward the AGN that is optically thin in the ultraviolet. This is consistent with the weak UV continuum. Rather, we note that the \civ\ line is a resonance line, so it is likely the emission is scattered continuum emission, where the continuum is produced by the central AGN at very small scales (sub-parsec). This implies that UV radiation from the AGN is escaping in a direction not along the line of sight to us, so that the covering factor $f_c$ of the outflow is less than unity.

In most quasars, the \civ\ line luminosity is a percent or so of the bolometric luminosity \cite{Baldwin77}. The observed flux of the CIV line relative to the bolometric luminosity is
\be
\frac{f_{\rm CIV}}{f_{\rm bol}}\approx 10^{-4}.
\ee
Thus, the product of the scattering optical depth times the covering factor of the \civ\ emitting gas we see is $f_{\rm cover,\ CIV}\times\tau_{\rm scatter}\approx 0.01$. If we take $f_{\rm cover,\ CIV}\approx0.1$ as inferred from the mid-IR emission models, the scattering optical depth is $\tau_{\rm scatter}\approx0.1$.

\begin{table}[!th]
    \centering
    \begin{tabular}{c|c|c}
        \hline 
        Parameter    & value         & unit\\
        \hline
        \hline
                & Dusty Nuclear Outflow &\\
                \hline
        $N_H$ & 4.5$\times10^{22}$ & cm$^{-2}$ \\
        V$\rm_{out}$  & $5250^{+276}_{-339}$    & \kms \\
        $R$ & 10-500 & pc \\
        $\dot{M}$ & 300-16,000 & \myr \\
        log10($\dot{E}_{outflow}$) &  45.3-47.2& $\ergs$\\
        log10($\dot{P}_{outflow}$) & 37-38.7 & g cm s$^{-2}$  \\
        $\frac{\dot{E}_{outflow}}{L_{bol}}$    & 0.2-10 &  \%       \\
        $\frac{\dot{P}_{outflow}}{\dot{P}_{AGN}}$ & 0.2-12   &      \\
                
                \hline
                 & Ionized Nuclear Outflow &\\
                 \hline
        logL(\ha)     & 43.65$\pm$0.04 & $\ergs$ \\
        W$_{90}$      &   11011 & \kms \\
        V$\rm_{out}$  & 9961    & \kms \\
        $R$ & $<$1200 & pc \\
        M$_{ionized}$ & $(3\pm1)\times10^{5}$ ($ \frac{\rm 9\times10^5 cm^{-3}}{n_{e}}$) & \msun \\
        $\dot{M}$    &  300 $\pm$50 ($ \frac{\rm 9\times10^5 cm^{-3}}{n_{e}}$)  & \myr \\
        log10($\dot{P}_{outflow}$) &  37.3$\pm$0.2                                         & g cm s$^{-2}$ \\
        log10($\dot{E}_{outflow}$) & 46$\pm$0.2                                       & $\ergs$       \\
        $\frac{\dot{E}_{outflow}}{L_{bol}}$    & 0.7$\pm$0.3 ($ \frac{\rm 9\times10^5 cm^{-3}}{n_{e}}$)       &                    \%       \\
        $\frac{\dot{P}_{outflow}}{\dot{P}_{AGN}}$ & 0.5$\pm$0.2 ($ \frac{\rm 9\times10^5 cm^{-3}}{n_{e}}$)      &                \\

    \end{tabular}
    \caption{Measured properties of the kinematics, dynamics, and energetics of the dusty and ionized nuclear outflows in W2246-0526.}
    \label{tab:nuclear_outflow_properties}
\end{table}

\pagebreak
\section{Radiation dust-driven winds}
\label{sec:radiation_driving}

In this section, we explore whether the dusty outflow can be driven through reprocessed infrared (as opposed to ultraviolet) radiation pressure alone for gas starting at the assumed circular velocity of 300 \kms\ at a launching radius (denoted by $R_0$) near the dust sublimation radius.

We use the Rosseland mean dust opacity from \cite{Semenov03}, $
\kappa_d=3\left(\frac{T}{100\,K}\right)^2\cm^2\,\g^{-1}$ for $T\le 100\ \K$, and,
\be
\kappa_d\approx6\,\cm^2\,\g^{-1}
\ee
for $200\,\K<T<600\,\K$ \cite{Pollack94}.

\noindent The spectral energy distribution (SED) for W2246-0526 is the best fit with two dust temperature models \cite{Fan16b,Tsai18}. The bulk of the luminosity is emitted by hot dust ($T\sim500-1000\,K$), while the remaining 15\% of the luminosity (see their table 3, \cite{Fan16b}) is emitted by dust with a temperature of $95\,\K$. The latter is dust heated by the quasar on larger scales up to 8 kpc \cite{Fernandez-Aranda24}, so we do not include the associated luminosity in the estimate of the radiation force acting on the gas in the inner regions ($< 1$ kpc) of the galaxy. In \cite{Fan16b}, the luminosity of the hot dust is computed by fitting a dusty torus SED and a second gray body model for the colder dust. We reconfirm that about 85\% of the total infrared radiation is emitted by the hot dust emission by directly integrating the rest-frame 1\micron\ - 600\micron\ SED and subtracting the contribution from $\sim100$K dust emission measured in \cite{Fernandez-Aranda24}.


Accordingly, we use an effective luminosity of
\be
L=f_{\rm mIR}*L_{\rm bol}=1.1\times10^{48}\,\rm erg~s^{-1},
\ee
where $f_{\rm mIR}=0.85$ is the fraction of the bolometric luminosity that emerges in the mid-IR, as defined by \cite{Fan16b}.

The gas column density inferred from the 3.4\micron\ absorption is
\be
N_H=4.5\times10^{22}\,\cm^{-2}
\ee
or in terms of the mass column density,
\be
\Sigma_H=7.5\times10^{-2} \g\ \cm^{-2}
\ee

%


Then the Rosseland mean absorption optical depth due to dust alone is greater than $\tau_d=\kappa_d\Sigma_H\approx0.4$, while the visual extinction is $A\rm_v\approx 23$.

The mean density of the mid-IR detected outflow is
\be
\bar \rho\approx \Sigma/R_0\approx2.4\times10^{-21}
\left(\frac{R_0}{10\,\pc}\right)\,\g\,\cm^{-3},
\ee

where we have assumed that the launch radius of the outflow $R_0\approx 10\,\pc$, just outside the dust sublimation radius. The corresponding number density is $\bar n_H\approx 1.5\times10^3\,\cm^{-3}$.

\subsection{Dynamics of a dust-driven wind}\label{sec: dynamics}
The radiation from the disk surrounding the central black hole is near, or possibly above, the Eddington limit \cite{Tsai18}. Since the Rosseland mean opacity to the (assumed reprocessed from UV to IR) quasar luminosity is a factor of 16 larger than the electron scattering opacity (of a fully ionized gas), the dusty gas surrounding the nucleus experiences a super Eddington luminosity and is accelerated outward.

In this section, we work out three relevant accelerations: the first due to radiation from the central AGN acting on dusty gas, the second due to the gravity from the central black hole, and the third from the galactic potential. As the gas flows outward, the radiative acceleration drops off at the same rate as that of the black hole gravity, but both fall more rapidly than the total acceleration due to gravity. When the magnitude of the gravitational acceleration equals the radiative acceleration, the wind begins to slow; we denote the radius at which this occurs by $R_{\rm decel}$.

The radiation force on the dust is
\be
\begin{split}
F_{\rm rad}&=f_c\tau_d f_{\rm mIR}\frac{L}{c}\\ 
&=1.7\times10^{37}
 f_c\,
\left(\frac{\kappa_d}{6\,\cm^2\,\g^{-1}}\right)
\left(\frac{\Sigma_H}{7.5\times10^{-2}\,\g\,\cm^{-2}}\right)
\left(\frac{f_{\rm mIR}}{0.85}\right)
\left(\frac{L_{\rm bol}}{1.35\times10^{48}\,\rm erg~s^{-1}}\right)
{\rm dyne},
\end{split}
\ee
where $ f_c$ is the covering factor of the dusty gas.

The mass of gas in the outflow is

\be
M_H=4\pi f_c R^2\Sigma_H=9\times10^{38} f_c
\left(\frac{R}{10\,\pc}\right)^2
\left(\frac{\Sigma_H}{7.5\times10^{-2}\,\g\,\cm^{-2}}\right)
\,\g
\ee

or $4.4\times10^5\,M_\odot$.

The acceleration of the gas, assuming the gas and dust are well coupled due to hydrodynamic and drag forces, is

\be \label{eqn: a rad}
a_{\rm rad}=\frac{F_{\rm rad}}{M_H}=\frac{\kappa_d}{4\pi R^2}\frac{f_{\rm mIR}L_{\rm bol}}{c}
\approx1.9\times10^{-2}
\left(\frac{\kappa_d}{6\,\cm^2\,\g^{-1}}\right)
\left(\frac{R}{10\,\pc}\right)^{-2}\,\cm\,\s^{-2}.
\ee

The inward acceleration due to the black hole is

\be \label{eqn: a BH}
a_{\rm BH}=\frac{GM_{\rm BH}}{R^2}=5.5\times10^{-4}
\left(\frac{M_{\rm BH}}{3.9\times10^{9}M_\odot}\right)
\left(\frac{R}{10\,\pc}\right)^{-2}\,\cm\,\s^{-2}.
\ee

It is worth noting that $a_{\rm rad}$ is about 35 times higher than $a_{\rm BH}$.

Since both these accelerations scale as $1/R^2$, it is convenient to define the ``Quasar acceleration'':
\be
a_Q(R)\equiv a_{\rm rad}-a_{\rm BH}=a_{\rm Q}(R_0)\left(\frac{R}{R_0}\right)^{-2},
\ee
where
\be
a_{\rm Q}(R_0)\equiv \frac{\kappa_d}{4\pi R_0^2}\frac{f_{\rm mIR}L_{\rm bol}}{c}-\frac{GM_{\rm BH}}{R_0^2}.
\ee

We note that if the luminosity is close to the Eddington luminosity of the central black hole, and the reprocessed radiation temperature is near $500\,\K$, then $a_{\rm Q}\approx a_{\rm rad}$.

These accelerations should be compared to the acceleration due to the gravity of the galaxy, 

\be \label{eqn: a gal}
a_{\rm gal}= \frac{v_c^2}{R}\approx2.9\times10^{-5}
\left(\frac{R_0}{R}\right)
\left(\frac{10\,\pc}{R_0}\right)
\left(\frac{v_c}{\rm 300\,km~s^{-1}}\right)^2
\cm\,\s^{-2},
\ee

where $v_c$ is the circular velocity of the galaxy, and we have assumed that $v_c$ is constant, independent of $R$.

While the galactic gravity is negligible compared to either the radiative acceleration or that of the central black hole near the dust sublimation radius, its dynamical effect relative to the radiative acceleration increases rapidly with increasing $R$. The inward and outward accelerations are equal at a radius given by $a_{\rm Q}(R)=a_{\rm gal}(R)$, or
\be\label{eqn: R decel}
R_{\rm decel}=\frac{\kappa_{\rm d}}{4\pi}\frac{f_{\rm mIR}L_{\rm bol}}{cv_c^2}-\frac{GM_{\rm BH}}{v_c^2}
\approx4.7
\left(\frac{300\,\rm km~s^{-1}}{v_c}\right)^2 \rm kpc.
\ee
Radiation cannot launch a wind at or outside this radius. We note that it is likely that $v_c$ is smaller than our assumed value at small radii (inside a few hundred parsecs), but we assume that $v_c$ reaches its asymptotic value by a kiloparsec or so since the galaxy is known to be very compact.

For winds launched at $R_0<R_{\rm decel}$, the gas first accelerates very rapidly, coasts for a while, then very slowly decelerates for $R>R_{\rm decel}$. Since the galaxy's gravity is relatively weak, the wind maintains a velocity close to $v_{\rm max}$ out to radii much larger than $R_{\rm decel}$, unless some new process, e.g., running into the interstellar medium or circumgalactic medium of the host galaxy, becomes relevant. 

The evolution of the wind is given by
\be
\frac{dv}{dt}=a_{\rm Q}(R)-a_{\rm gal}(R)
\ee
and
\be
v=\frac{dR}{dt}.
\ee

Combining, we find
\be \label{eqn: dvdr}
v(R)\frac{dv}{dR}=a_{\rm Q}(R)-a_{\rm gal}(R).
\ee
At small radii, where $a_{\rm Q}\gg a_{\rm gal}$, this reduces to

\be
\int vdv=a_{\rm Q}(R_0)\int \left(\frac{R_0}{R}\right)^2dR
\ee

Integrating, we find
\be \label{eqn: velocity}
v(R;R_0) = v_Q(R_0)v_N(R;R_0)
\ee
where we have introduce the maximum wind velocity

\be
v_{\rm Q}=\sqrt{a_{\rm Q}(R_0)R_0}=
\sqrt{\frac{\kappa_d}{2\pi R_0}\frac{f_{\rm mIR} L_{\rm bol}}{c}-\frac{GM_{\rm BH}}{R_0}}\approx10,830
\left(\frac{10\,\rm pc}{R_0}\right)^{1/2}\,\rm km~s^{-1},
\ee

and the normalized velocity
\be
v_N(R;R_0)\equiv\left(
\left[1-\frac{R_0}{R}\right]-\frac{2v_c^2}{v_Q^2(R_0)}\ln\frac{R}{R_0}
\right)^{1/2}.
\ee

Since $v_c\ll v_Q$, we will use the following approximation, good to about a tenth of a percent:
\be
v(R)\approx\sqrt{2a_{\rm Q}(R_0)R_0}\sqrt{1-\frac{R_0}{R}}.
\ee

A useful alternative expression for the terminal velocity can be given in terms of the Eddington ratio
\be
\Gamma\equiv \frac{L_{\rm bol}}{L_{\rm Edd}},
\ee
with
\be
L_{\rm Edd}\equiv\frac{4\pi GM_{\rm BH}c}{\kappa_{\rm es}},
\ee
where $\kappa_{\rm es}\approx 0.38\,\cm^2\,\g^{-1}$ for ionized gas. Using this,

\be
v_Q(R_0)=\sqrt{\frac{\kappa_d}{2\pi R_0}\frac{f_{mIR}L_{\rm bol}}{c}
  \left[1-\frac{1}{2f_{mIR}}\frac{\kappa_{\rm es}}{\kappa_d}\frac{1}{\Gamma}\right]}.
\ee

\begin{figure}
  \includegraphics{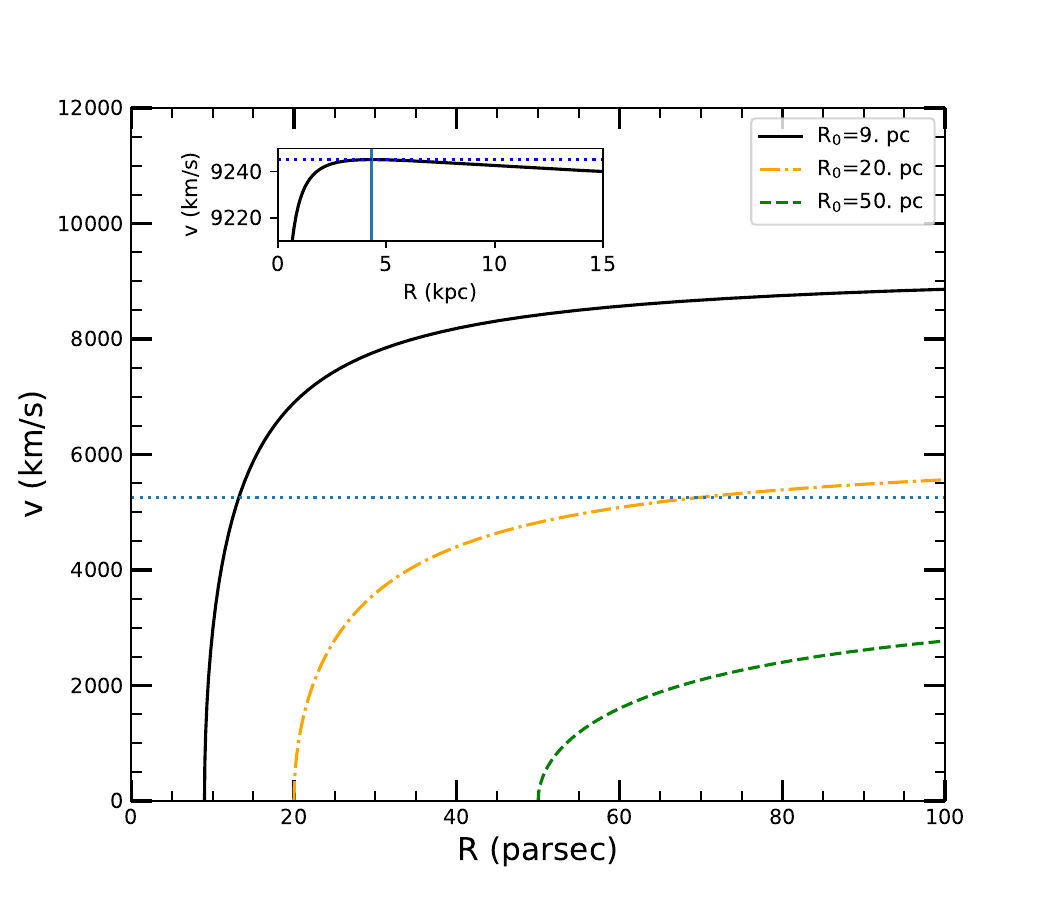}
  \caption{The run of velocity for dust-driven winds launched at radii of $9\,\pc$ (black line), $20\,\pc$ (orange dash-dot line) and $50\,\pc$ (green dashed line). The horizontal dotted line is at the outflow velocity detected in W2246-0526. The maximum line optical depth is set by a trade off between the velocity shear and the gas density. A velocity shear (or acceleration) leads to a small optical depth, while a large density leads to a large optical depth. The density is inversely proportional to the velocity and to the square of the radius from the black hole. As a result, we expect that the maximum optical depth occurs at an intermediate velocity. The inset shows the  $R_0=9\,\pc$ velocity out to $15\,\kpc$ illustrating the decrease in velocity beyond $R_{\rm decel}$ from eqn. (\ref{eqn: R decel}) denoted by the vertical blue line.\label{fig: velocity}}
\end{figure}

Figure \ref{fig: velocity} shows the run of velocity for three different launch radii, namely $R_0=9$, $R_0=20$, and $R_0=50$ parsecs. If the wind is launched at $R_0=500~\pc$,
\be
v_{\rm Q}(R_0=500\,\pc)\approx1,000\,\rm km~s^{-1},
\ee
well below the observed outflow velocity. This suggests that the outflow starts at or near the dust sublimation radius ($10$ pc), and well inside $R=500$ pc. The orange dashed-dot line in the Figure \ref{fig: velocity} shows a wind with a terminal velocity slightly above that observed in W2246-0526. The acceleration from rest to the near plateau value happens very rapidly, only taking on the order of a thousand years. However, during this acceleration phase, the gas and dust are spread over a broad range of velocities. At any given velocity, this translates to a lower optical depth, hence a weaker absorption signal. At radii of 2-3 times the original driving radius, the velocity range becomes significantly lower, meaning the majority of the dust and gas are moving at similar velocities with a much smaller range compared to the acceleration phase. This translates to a larger build-up in optical depth, making the signal easier to detect in absorption. The peak of the optical depth likely occurs near the turnover of the velocity (2-3 $\times R_{0}$). This further indicates that the signal that is more easily detectable is expected to have an intrinsically narrow velocity dispersion. This is likely the reason why we do not see any significant broadening of the aromatic or aliphatic absorption profiles relative to what is observed in the Milky Way.


\subsection{The mass loss rate, momentum loss rate, and kinetic luminosity of the dusty outflow} \label{sec: mass}

The continuity equation in spherical geometry is
\be
\frac{\partial\rho}{\partial t} +\frac{1}{r^2}\frac{\partial r^2\rho v}{\partial r}=0,
\ee
while the momentum equation is given by using equations (\ref{eqn: a rad}), (\ref{eqn: a BH}), and (\ref{eqn: a gal}) in equation (\ref{eqn: dvdr}).

In the body of the wind, i.e., above $\tau\approx1$, the time derivative is negligible, so the continuity equation reduces to the statement that $r^2\rho v=const.$, leading to the expression of the wind mass loss rate:
\be \label{eqn: mdot}
\dot M_w = 4\pi f_c R^2\rho(R)v(R; R_0).
\ee

\begin{figure}
  \includegraphics{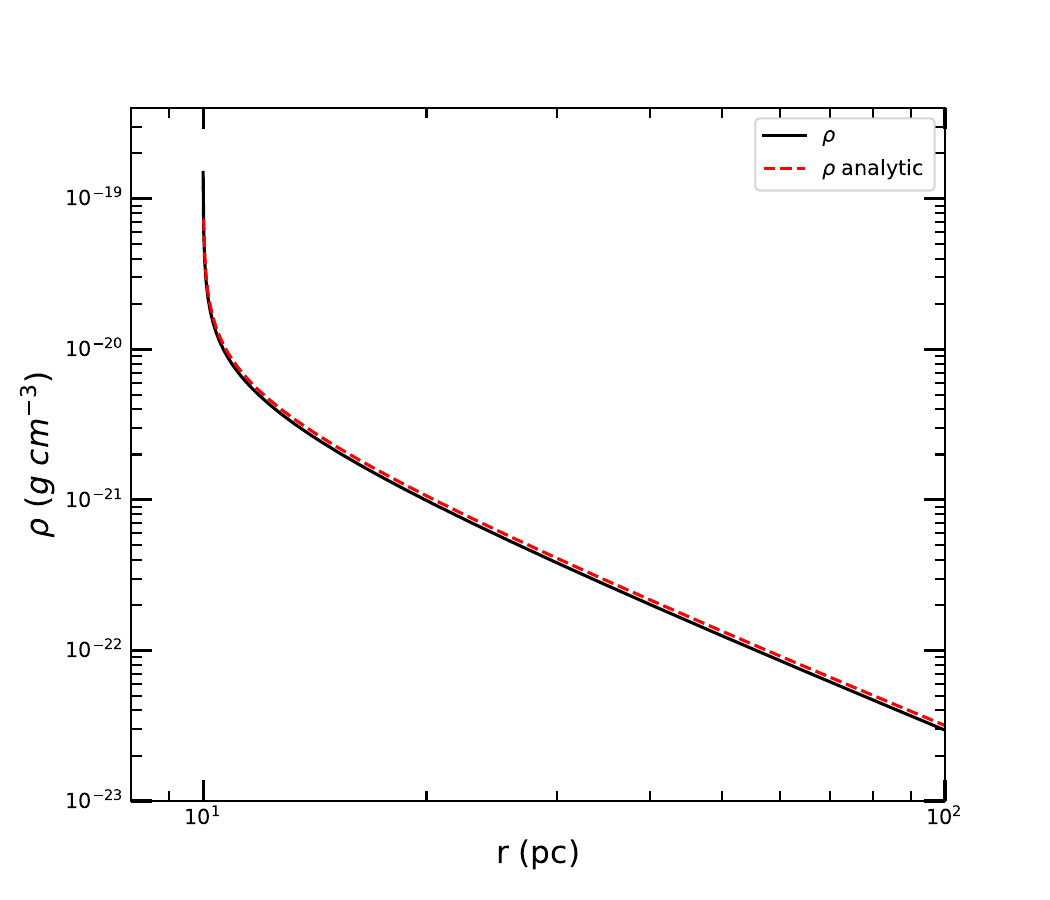}
  \caption{The run of density for a dust-driven wind launched at  $R_0=10\pc$. The solid black line is from a numerical integration, while the dashed red line is from equation \ref{eqn: rho thin} \label{fig: density}}
\end{figure}

The density in the wind is given by

\be \label{eqn: rho thin}
\rho_w(R)=\frac{f_{mIR}L_{\rm bol}}{4\pi R^2 v_Q^2 c}\,\frac{1}{v_N(R;R_0)}.
\ee

This expression for the density is only valid above the hydrostatic region, roughly corresponding to $\tau~\gtrsim$1.
Figure \ref{fig: density} shows the numerically calculated run of density in the wind.

\begin{figure}
\includegraphics[]{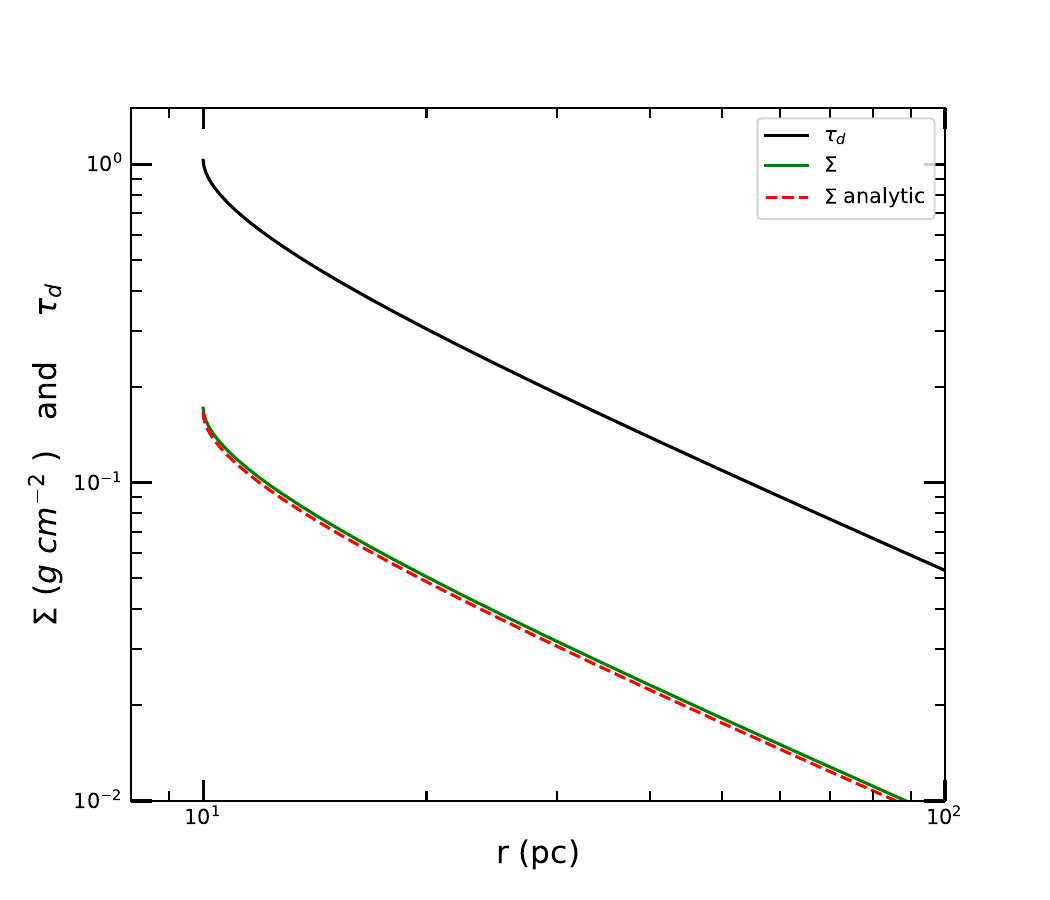}
\caption{The run of column density $\Sigma(r)$ with radius from equation \ref{eqn: Sigma thin} (dashed red line) and from numerical integration (solid green line), and the mid-IR optical depth $\tau(r)$ (solid black line), all for a dust driven wind launched at $R_0=10\pc$. \label{fig: Sigma}}
\end{figure}

The column density is
\be \label{eqn: Sigma thin}
\Sigma_w(R)\approx\frac{1}{2\pi R_0 v_Q^2}\frac{f_{mIR}\lbol} {c}\sqrt{1-\frac{R_0}{R}}
\approx 0.2 \g\,\cm^{-2},
\ee
where the prefactor is independent of $\lbol$ and $R_0$. We have neglected terms proportional to $(v_c/v_Q)^2$; this approximation is also accurate to about a percent. 

Figure (\ref{fig: Sigma}) shows the run of column density and optical depth $\tau$ in the wind. We remind the reader that Equations (\ref{eqn: mdot}-\ref{eqn: Sigma thin}) apply only for $\tau\gtrapprox1$. 

For the model shown, the outflow rate in the wind is $\approx 600$ \myr, while the kinetic luminosity has a maximum value $L_w\approx2\times10^{46}\,{\rm erg}\,{\rm s}^{-1}$, which equates to a kinetic luminosity of $\approx 1$\% of the quasar bolometric luminosity.

Our calculations show that radiation pressure on ordinary dust grains alone is capable of driving the observed 5250 \kms\ dusty outflow in the W2246-0526 system with incredibly rapid acceleration. We have in hand calculations indicating that the best fit to the observed absorption feature arises in a wind launched at $R_0=10\pc$, with a terminal velocity exceeding $10,000\,{\rm km}\,{\rm s}^{-1}$. In that model, the bulk of the opacity arises from radii of $20-30\,\pc$, so that the deepest absorption occurs around $5200\,{\rm km}\,{\rm s}^{-1}$





\end{document}